\documentclass[aps,prb,preprint,superscriptaddress,showpacs]{revtex4-1}

\usepackage{graphicx}
\usepackage{dcolumn}
\usepackage{amsmath,bm}
\usepackage{color}
\usepackage{multirow}

\begin{document}

\newpage

\title{\textit{Ab initio} Study of Aspirin Adsorption on Single-walled
Carbon and Carbon Nitride Nanotubes}

\author{Yongju Lee}
\affiliation{Department of Physics and
             Research Institute for Basic Sciences,
             Kyung Hee University, Seoul, 02447, Korea}

\author{Dae-Gyeon Kwon}
\altaffiliation[Current address: ]{LG Chem Research Park, Daejeon 34122, Korea}
\affiliation{Department of Physics and
             Research Institute for Basic Sciences,
             Kyung Hee University, Seoul, 02447, Korea}

\author{Gunn Kim}
\email[Corresponding author. E-mail: ]{gunnkim@sejong.ac.kr}
\affiliation{Department of Physics \& Astronomy and
             Graphene Research Institute,  
             Sejong University, Seoul, Korea, 05006, Korea}

\author{Young-Kyun Kwon}
\email[Corresponding author. E-mail: ]{ykkwon@khu.ac.kr}
\affiliation{Department of Physics and
             Research Institute for Basic Sciences,
             Kyung Hee University, Seoul, 02447, Korea}

\date{\today}

\begin{abstract}
We use \textit{ab intio} density functional theory to investigate the
adsorption properties of acetylsalicylic acid or aspirin on a $(10,0)$
carbon nanotube (CNT) and a $(8,0)$ triazine-based graphitic carbon
nitride nanotube (CNNT). It is found that an aspirin molecule binds
stronger to the CNNT with its adsorption energy of $0.67$~eV than to
the CNT with $0.51$~eV. The stronger adsorption energy on the CNNT is
ascribed to the high reactivity of its N atoms with high electron
affinity. The CNNT exhibits local electric dipole moments,
which cause strong charge redistribution in the aspirin molecule
adsorbed on the CNNT than on the CNT. We also explore the influence
of an external electric field on the adsorption properties of aspirin
on these nanotubes by examining the modifications in their electronic
band structures, partial densities of states, and charge
distributions. It is found that an electric field applied along
a particular direction induces aspirin molecular states in the in-gap
region of the CNNT implying a potential application of aspirin
detection.

%
\end{abstract}


%



\maketitle



\section{Introduction}
\label{Introduction}

Acetylsalicylic acid (ASA), also known as aspirin, has been one of the
most widely used medications in the world due to its well-known
effects of reducing fevers and relieving aches and pains. Aspirin has
been used to help prevent heart attacks, strokes, blood clot
formation and suppression of prostaglandin owing to its antiplatelet
effect of decreasing platelet aggregation and inhibiting thrombus
formation.~\cite{Cheng2007} It has also been reported that aspirin has
an anticancer effect~\cite{Ferreira1973} and a precautionary effect on
stroke~\cite{Lewis1983}. Despite of such good effects, however, it has
been warned that such a famous non-prescription medicine may provoke
some adverse effects and thus people should avoid its substance (drug)
abuse and misuse. Because aspirin has an effect on the stomach 
lining, it is recommended that people with gastroenteric disorders such as 
gastritis and peptic ulcers take medical advice before using aspirin. When 
aspirin is taken with alcohol, stomach bleeding may occur even to healthy 
people. Therefore, it is important to measure the amount of ASA molecule in solution or 
in human body.

Carbon nanotubes (CNTs) are thin and long macromolecules of graphitic carbon 
that have attracted huge academic and industrial interest because of unique 
physical and chemical properties~\cite{Chico1996, Tans1998, Kwon1999, Yao1999, 
Sanvito1999, Berber2000, Zhou2000, Lee2002, Kim2005, Charlier2007, Choi2008}. 
One of the attractive characteristics of the CNT is a very large adsorption 
surface area~\cite{Long2001a, Cinke2002, Yin2000} comparable to that of 
carbon-based adsorbents (such as activated carbon) used commercially. Several 
binding properties of molecules onto the CNT have been also studied 
theoretically and experimentally~\cite{Dillon1997, Chambers1998, Long2001a, 
Peng2003, Fagan2004, Fagan2006, Tournus2005} for its applications. Using the 
property of the large adsorption surface areas, CNT can be used to filter or to 
detect molecules. Since some gas molecules such as ammonia 
(NH$_{3}$)~\cite{Kong2000}, nitrogen dioxide (NO$_{2}$)~\cite{Kong2000}, 
alcohol~\cite{Song2008}, and other molecules~\cite{Collins2000, Bradley2003, 
Yu2000, Burt2005} were reported to be detectable by these devices, a wide range 
of molecules have been studied for chemical sensing of CNT. Due to the 
modification of intrinsic electronic structures of semiconducting CNT by 
adsorption, the type and concentration of target molecules may be detected.

Carbon nitride compounds have been studied in various area such as electronic 
devices, humidity sensor, and coatings because of their electronic and chemical 
properties~\cite{Zhang2009, Li1995, Zambov2000}. They have various structures, 
depending on their carbon to nitrogen atomic ratio and arrangement. In this 
study, we focus on a graphitic carbon nitride (g-C$_{3}$N$_{4}$) nanotube 
(CNNT), triazine-based form. Recent studies report g-C$_{3}$N$_{4}$ nanotube with 
respect to synthesis and first-principles calculations~\cite{Cao2004, Guo2004, 
Li2007, Pan2011}. Various adsorption properties are expected because of the 
unique porous structure.

In this paper, we present a first-principles study of binding properties of ASA 
on the bare CNT and the bare CNNT. ASA binds to pristine CNT (CNNT) with binding 
energy of $0.51$~eV ($0.67$~eV), and no practical charge transfer takes place. 
According to our analysis on the electronic structure, 
CNT with ASA adsorbates does not show significant characteristics, but 
CNNT with ASA adsorbates show noticeable effects. Because of the 
structure of the CNNT, local electric dipole moments occur and interrupt 
exchange of electrons between the ASA molecule and the CNNT. 
Finally we discuss homogeneous external electric field (E-field) effects on the 
CNNT with ASA adsorbates.
The response of nanotubes to the E-field is attractive for the application to 
electric devices.

\section{Computational method}
\label{Computational}

Using the first-principles calculations based on density functional theory 
(DFT)~\cite{Hohenberg1964, Kohn1965}, we examined the CNT and the CNNT with 
respect to the adsorption of ASA. The electronic wavefunctions were expanded 
into plane waves up to a cutoff energy of $450$~eV, and the ion-electron 
interactions were described using the projector augmented wave method 
implemented in the Vienna \textit{ab initio} simulation package (VASP)~\cite{ 
Kresse1993, Kresse1996}, within the generalized gradient approximation (GGA)
method~\cite{Perdew1996}. To better describe interaction between ASA and 
nanotubes, we considered van der Waals interaction using Grimme's method 
(DFT-D2)~\cite{Grimme2006}. All the model structures were relaxed until the 
Hellmann--Feynman forces were smaller than $0.03$~eV/{\AA}. We chose the 
$(10,0)$ CNT and the $(8,0)$ CNNT as host materials for molecular 
adsorption. For the calculation of the ASA molecule in vacuum, we used a cubic 
supercell with a length of 30~{\AA}. The $\Gamma$-point were used in 
calculations of the isolated ASA molecule. The lengths of the supercell along 
the tube axis were $16.05$~{\AA} for the CNT, and $15.72$~{\AA} for the CNNT, 
respectively. We used one $k$ point (the $\Gamma$-point) for the nanotubes in the 
geometry optimization, and $1\times 1 \times 10$ Monkhorst-
Pack~\cite{Monkhorst1976} $k$ point sampling in the electronic structure 
calculations. For each adsorption configuration, we calculated binding energy 
($E_{b}$) defined by Eq.~\ref{equation1}. $E$[tube+ASA] is the total energy of 
the CNT and the CNNT with an ASA molecule, and $E$[tube] is the total energy of 
the two nanotubes without any ASA molecule, respectively. $E$[ASA] also shows 
the energy of the isolated ASA molecule.

\begin{align}\label{equation1}
\textit{E}_{b} = \textit{E}[\rm{tube}] + \textit{E}[\rm{ASA}] - \textit{E}[\rm{tube+ASA}]
\end{align}

\section{Results and discussion}
\label{Results}

\begin{figure}[t]
\includegraphics[width=1.0\columnwidth]{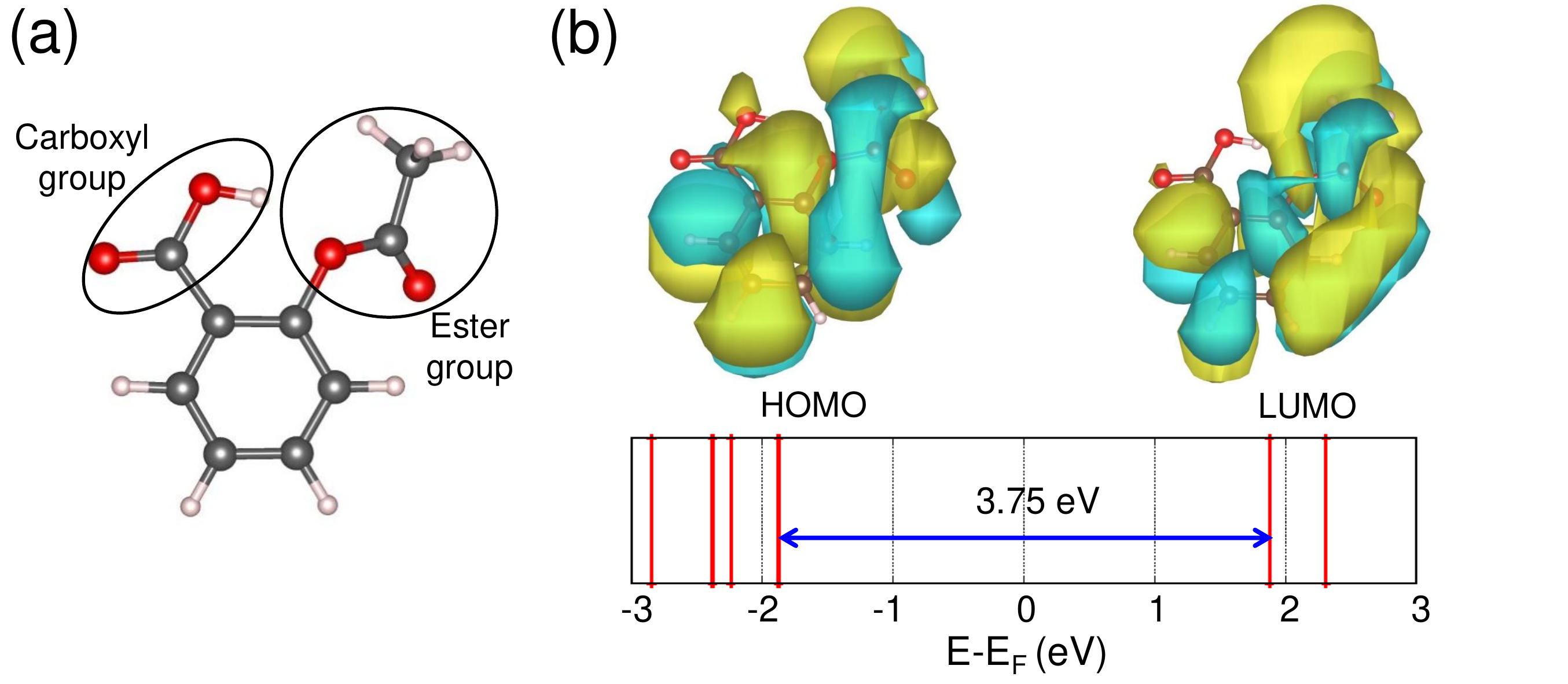}
\caption{(a) Model structure of an ASA molecule. (b) Electronic energy levels of 
an ASA molecule with wavefunctions of its HOMO and LUMO. The HOMO--LUMO gap is 
$3.75$~eV.
\label{asaFig1}}
\end{figure}

\begin{figure}
\includegraphics[width=1.0\columnwidth]{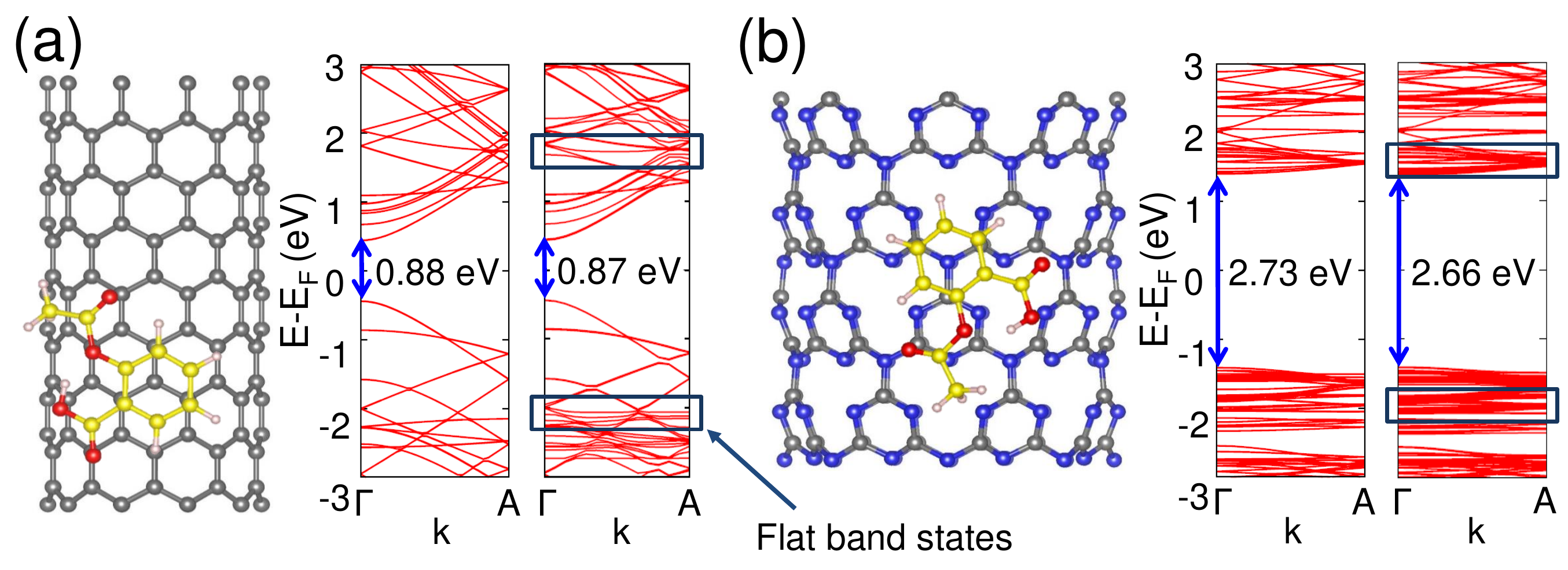}
\caption{Electronic property of ASA adsorbed on (a) the CNT and (b) the CNNT. The most stable adsorption sites among possible configurations (left) and band structures of the bare nanotube (middle) and ASA on nanotube (right). The flat bands in the rectangles originate from the ASA molecule
on the tube.
\label{asaFig2}}
\end{figure}

Fig.~\ref{asaFig1}(a) shows the most stable structure of an ASA molecule 
(C$_{9}$H$_{8}$O$_{4}$) in vacuum. It consists of an aromatic ring, ester, and 
carboxylic acid. Fig.~\ref{asaFig1}(b) and (c) show the highest occupied 
molecular orbital (HOMO) and the lowest unoccupied molecular orbital (LUMO), 
respectively. Its HOMO--LUMO gap is $3.75$~eV. In the HOMO and the LUMO of ASA, 
electronic densities are distributed over the entire molecule, as shown in 
Fig.~\ref{asaFig1}(b). We selected the two types of nanotubes to investigate 
their application for detecting ASA molecules: the $(10,0)$ CNT and the $(8,0)$ 
g-C$_{3}$N$_{4}$ nanotube. As shown in Fig.~\ref{asaFig2}(a) and (b), the bare 
$(10,0)$ CNT has a band gap of $0.88$~eV, and the bare $(8,0)$ CNNT has a band gap 
of $2.55$~eV.

When an ASA molecule is adsorbed on nanotubes, many possible configurations can 
be considered. 
We chose only $\pi-\pi$ stacking configurations, and calculated the total energy 
of each configuration 
to obtain the most stable adsorption site.  
For visual clarity, the carbon atoms of the ASA molecule is shown in yellow, 
while those of the CNT and the CNNT is represented in black 
(Fig.~\ref{asaFig2}). For the binding of ASA on the $(10,0)$ CNT, the most 
stable 
configuration resembles Bernal stacking (AB stacking) of graphite, as seen in 
Fig.~\ref{asaFig2}(a). 
Its electronic band structure shows no clear change near the Fermi level 
compared with that of the bare CNT, 
even if flat bands originating from the ASA molecule are shown around $-2$~eV 
and $+2$~eV. 
Fig.~\ref{asaFig2}(b) shows the most stable adsorption configuration of the CNNT 
with ASA adsorbates. Nitrogen atoms located in the benzene ring of ASA and the 
carboxyl group of ASA is placed just above pore sites of the CNNT. The band gap 
slightly decreases by $0.07$~eV. 
Since $\pi-\pi$ hybridization is the main origin of the molecular binding, 
noticeable charge transfer does not occur.
Consequently any molecular state from ASA is not shown in the band gap of the 
CNT or the CNNT.
We conclude that it is difficult to detect ASA using the CNT or the CNNT in the 
absence of external E-field.
Table~\ref{asatable1} summaries the adsorption energy and the distance between 
the ASA molecule and the tube wall for each adsorption configuration. We find 
that ASA binds more strongly to the CNNT than the CNT.

\begin{figure}
\includegraphics[width=0.9\columnwidth]{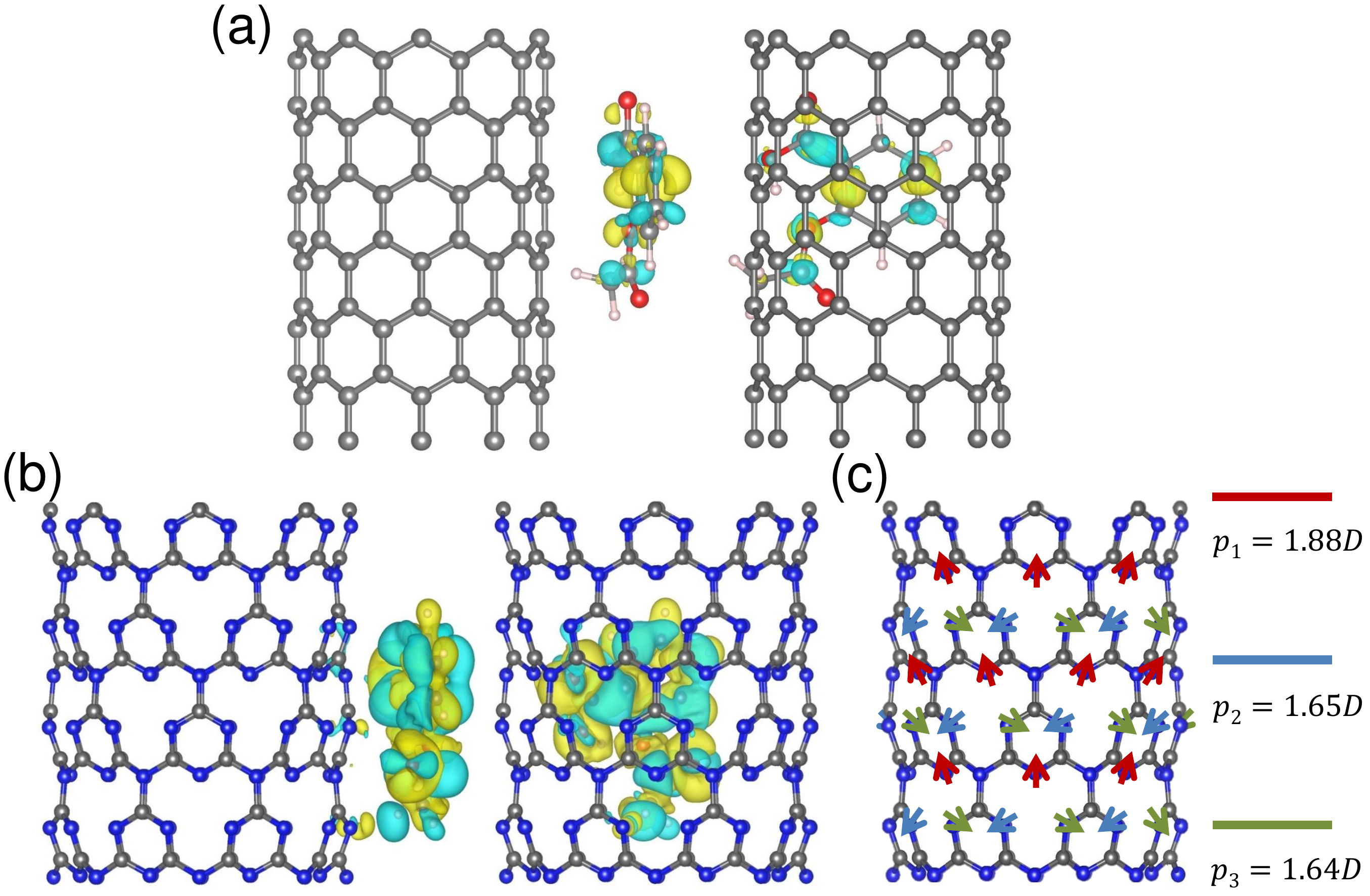}
\caption{(a) Charge difference of the ASA-adsorbed CNT and (b) charge difference 
of the ASA-adsorbed CNNT. (c) Directions and magnitudes of electrical dipole 
moments in the bare CNNT. Cyan and yellow colors represent 
electron accumulation and electron depletion, respectively. D is the unit of the 
electric dipole moment, debye.
\label{asaFig3}}
\end{figure}

To understand the difference in the adsorption energy between the CNT and the 
CNNT, we studied
the interaction between the ASA molecule and the CNT and CNNT in 
more detail.
Fig.~\ref{asaFig3}(a) and (b) shows a charge difference [$\rho({\rm tube+ASA}) - 
\rho({\rm tube}) - \rho({\rm ASA})$] plot of the $(10,0)$ CNT and $(8,0)$
CNNT with ASA, respectively. As mentioned above, there is no charge transfer between 
the ASA molecule and both nanotubes. Interestingly, we observe that the much bigger 
charge redistribution appears in the ASA molecule with CNNT than CNT. What makes 
the charge redistribution in ASA?
Because of the electronic configurations of the C and N atoms, the nanotube 
forms a buckled structure contrary to CNT's structure. Lone pair electrons, 
which are localized at the N atoms, result in buckling of the tube surface. 
In addition, we can take into account local electric dipole moments of the CNNT. 
The difference in the electron affinity of the C and N atoms causes the local 
electric dipole moments, which are obtained by Eq.~\ref{equation2}. Electric 
dipole moments of the CNNT are shown in Fig.~\ref{asaFig3}(c). As a result, 
local electric dipole moments together with buckling and vacancies bring about 
the charge redistribution of ASA.

\begin{table}[b]
\caption{
The binding energy and the distance between the adsorbed molecule and nanotubes. 
(C$_{3}$N$_{4}$ + ASA 2, 3 in Supplementary Information.)
\label{asatable1}}
\begin{ruledtabular}
\begin{tabular}{cccccccc}
Structure & Binding energy (eV) & Distance (\AA) \\ 
\hline
CNT + ASA & 0.54 & 2.59 \\
C$_{3}$N$_{4}$ + ASA 1 & 0.66 & 2.70 \\
C$_{3}$N$_{4}$ + ASA 2 & 0.68 & 2.68 \\
C$_{3}$N$_{4}$ + ASA 3 & 0.67 & 2.72 \\
\end{tabular}
\end{ruledtabular}
\end{table}
\begin{align}\label{equation2}
\int \rho(r)rdr+\sum_{i}Z_iez_i
\end{align}

Fig.~\ref{asaFig3}(c) depicts local electric dipole moments in the surface of 
the CNNT. 
Above all, the nitrogen atoms with red arrows are more buckled in a different 
direction from those with green and blue arrows.
Therefore, the dipole moments denoted by green and blue arrows have almost 
similar values, $1.64$~D and $1.65$~D, respectively. 
($1~{\rm D} \approx 3.336 \times 10^{-30}~ {\rm C \cdot m} \approx 0.2082$ 
e$\cdot$\AA), respectively, while those by red arrows are bigger ($1.88$~D). 
Because of the local electric dipole moments, the ASA molecule has dipole-dipole 
interaction with the CNNT. ASA in vacuum has an electric dipole 
moment of $1.51$~D, but ASA on the CNNT has a larger moment 
of $2.25$~D. If a simple dipole-dipole interaction ($V_{dd}$ in 
Eq.~\ref{equation3}) 
is applied with the aforementioned dipole moments at the ASA molecule and the 
CNNT, the interaction energy is about $-0.1$~eV. It explains the difference in 
the binding energy listed in Table~\ref{asatable1}. We conclude that the aspirin 
molecule adsorbs more strongly on the CNNT than the CNT owing to the dipole-dipole interaction.

\begin{align}\label{equation3}
V_{dd} = \frac{1}{4\pi\epsilon_{0}}\left[\frac{\vec{p}_{1}\cdot\vec{p}_{2}}{|\vec{r}_{1} - \vec{r}_{2}|^3}-\frac{3\{(\vec{r}_{1} - \vec{r}_{2})\cdot \vec{p}_{1}\}\{(\vec{r}_{1} - \vec{r}_{2})\cdot \vec{p}_{2}\}}{|\vec{r}_{1} - \vec{r}_{2}|^5}\right],
\end{align}
where $V_{dd}$ is a potential energy for dipole-dipole interaction, and $\vec{p}_i$ and $\vec{r}_i$ 
($i$ = 1, 2) are the $i$-th electric dipole moment and its position, respectively.

\begin{figure}
\includegraphics[width=1.0\columnwidth]{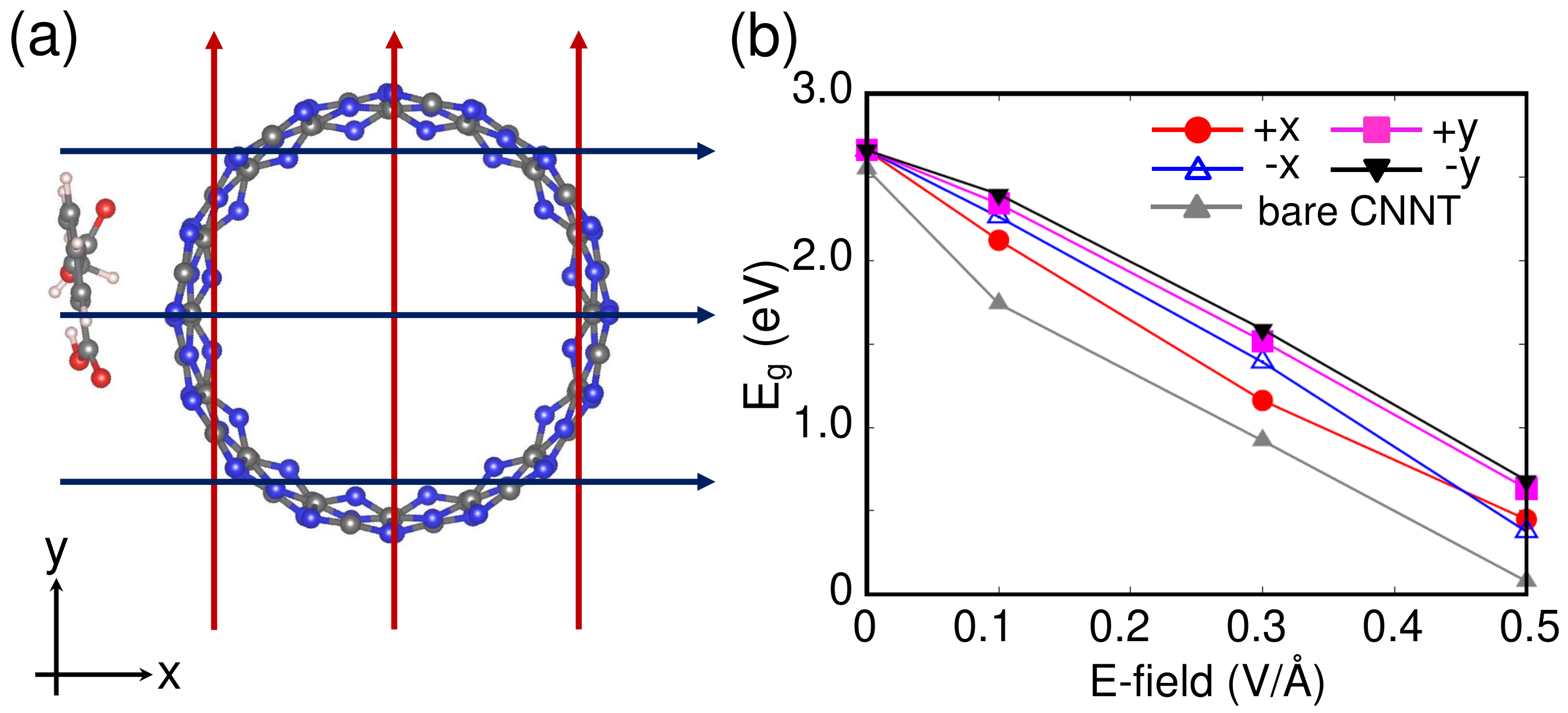}
\caption{(a) Directions of external electric field on the ASA-adsorbed 
CNNT. (b) Band gaps of bare CNNT and ASA-adsorbed CNNT in all 
directions of the external electric field. The $x$ directions in (a) cross both 
of the ASA molecule and the CNNT perpendicularly. The $y$ 
directions also cross CNNT perpendicularly, but ASA molecule 
horizontally. 
They tend to decrease according to intensity of electric field.
\label{asaFig4}}
\end{figure}

\begin{figure}
\includegraphics[width=1.0\columnwidth]{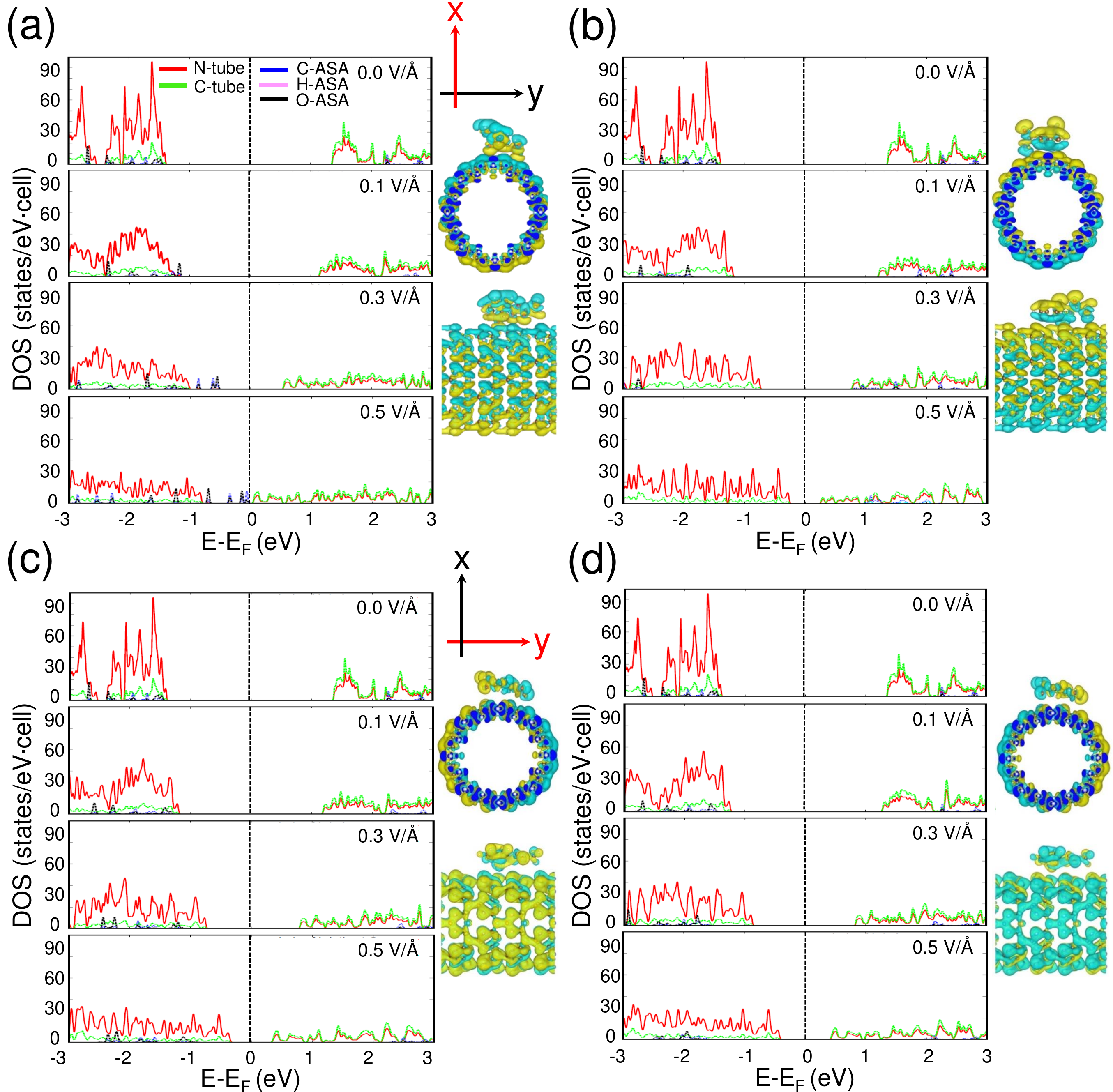}
\caption{(a-d) PDOS plots and electron density differences of the ASA-adsorbed 
CNNT under the external E-fields in $\pm x$ and $\pm y$ 
directions, respectively. The arrow means the direction of E-field. Charge 
transfer occurs due to the external E-field. Cyan and yellow colors represent 
electron accumulation and electron depletion, respectively.
\label{asaFig5}}
\end{figure}

To check whether the nanotube can be applied to a detecting device for ASA 
molecules, we checked effects of external E-field on electric property of the 
bare CNNT and ASA-adsorbed CNNT. 
In a single-gated field effect transistor, an E-field is generated by an applied 
gate bias between the gate and the nanotube. Our computational results reveal 
the difference between the electronic structure of the bare CNNT and 
ASA-adsorbed CNNT. Band gap of the bare CNNT is shown much smaller than 
ASA-adsorbed CNNT. Not only there are the difference of band gap between the 
bare CNNT and ASA-adsorbed CNNT, but also the different electronic property 
among four directions. The electronic structure of the ASA-adsorbed CNNT is not 
affected very much by the E-field perpendicular to the tube axis. In contrast, 
the ASA-adsorbed CNNT show remarkable characters in the band structure, depending on 
the direction of the transverse E-field, as shown Fig.~\ref{asaFig4}. We chose 
four directions of the E-field ($\pm x$ and $\pm y$), which were all 
perpendicular to the CNNT axis direction. The E-fields in $\pm x$ directions may 
enhance electronic coupling between the ASA molecule and the CNNT, whereas the 
fields in the $\pm y$ directions show the weak ASA-CNNT coupling. 
Fig.~\ref{asaFig4}(b) shows that the external E-field can modulate the 
electronic structure of the ASA-adsorbed CNNT. We also present partial densities 
of states (PDOS) of the ASA-adsorbed CNNT in Fig.~\ref{asaFig5}. Since the 
E-field is in the $+x$ direction and the CNNT has a higher electric potential 
than the ASA adsorbate, electrons are donated from adsorbed ASA molecule to one 
part of the CNNT. Fig.~\ref{asaFig5}(a) shows that the transverse E-field in the 
$+x$ direction makes localized states of the ASA molecule be upshifted, and move 
into the + region of the CNNT, as the E-field strength 
increases. Besides the conduction band of the tube also moves down to the Fermi 
level. On the other hand, the E-fields in $-x$, $+y$ and $-y$ directions have 
the similar tendency; Although the band gap decreases, the states originating 
from ASA do not appear as the in-gap states. In both $\pm y$ directions, 
external E-field does not cause strong ASA-CNNT coupling enhancement, which is related 
to the reflection symmetry. The band gap of the bare CNNT is also reduced under 
external E-field, but there is no localized states in the energy band gap in 
this case. In real situations, the CNNT would be coated with ASA molecules 
randomly, and thus any direction of the E-field makes the ASA states occur in 
the forbidden band. Such in-gap states could result in the scattering in the 
electron transport, which would be reflected in the current-voltage curve in 
experiment. Consequently, the C$_{3}$N$_{4}$ nanotube-based sensor device may 
detect the ASA molecules using the gate bias voltage (or external E-field). 

\section{Conclusion}
\label{Summary}

In summary, we have performed {\it ab initio} calculations within density 
CNNT. ASA is adsorbed more weakly on the CNT than the 
CNNT. Since CNNT have local electric dipole moments, it causes a stronger binding to 
ASA than the CNT owing to dipole-dipole interaction of the CNNT with the ASA 
molecule. When a homogeneous external electric field is introduced, the band gap of 
the CNNT decreases dramatically in the presence of the ASA adsorbates. 
Especially, when $+x$ direction is applied, in-gap states of aspirin molecule occur. 
Thus we expect that CNNT could be used in application to chemical sensors based on 
the field effect transistor.



\section*{Acknowledgments}

We gratefully acknowledge financial support from the Korean government
through National Research Foundation (NRF-2011-0016188). 
GK were supported by the Priority Research Center Program
(2010-0020207) and the Basic Science Research Program (2013R1A2009131) through 
NRF. 
Some portion of our computational work was done using the resources of the KISTI 
Supercomputing Center (KSC-2015-C3-023).

%
\bibliographystyle{apsrev}
\bibliography{asa.ref.bbl} 

\begin{thebibliography}{43}
\expandafter\ifx\csname natexlab\endcsname\relax\def\natexlab#1{#1}\fi
\expandafter\ifx\csname bibnamefont\endcsname\relax
  \def\bibnamefont#1{#1}\fi
\expandafter\ifx\csname bibfnamefont\endcsname\relax
  \def\bibfnamefont#1{#1}\fi
\expandafter\ifx\csname citenamefont\endcsname\relax
  \def\citenamefont#1{#1}\fi
\expandafter\ifx\csname url\endcsname\relax
  \def\url#1{\texttt{#1}}\fi
\expandafter\ifx\csname urlprefix\endcsname\relax\def\urlprefix{URL }\fi
\providecommand{\bibinfo}[2]{#2}
\providecommand{\eprint}[2][]{\url{#2}}

\bibitem[{\citenamefont{Cheng}(2007)}]{Cheng2007}
\bibinfo{author}{\bibfnamefont{T.~O.} \bibnamefont{Cheng}},
  \bibinfo{journal}{Texas Hear. Inst. J.} \textbf{\bibinfo{volume}{34}},
  \bibinfo{pages}{392} (\bibinfo{year}{2007}).

\bibitem[{\citenamefont{Ferreira et~al.}(1973)\citenamefont{Ferreira, Moncada,
  and Vane}}]{Ferreira1973}
\bibinfo{author}{\bibfnamefont{S.~H.} \bibnamefont{Ferreira}},
  \bibinfo{author}{\bibfnamefont{S.}~\bibnamefont{Moncada}}, \bibnamefont{and}
  \bibinfo{author}{\bibfnamefont{J.~R.} \bibnamefont{Vane}},
  \bibinfo{journal}{Br. J. Pharmacol.} \textbf{\bibinfo{volume}{49}},
  \bibinfo{pages}{86} (\bibinfo{year}{1973}).

\bibitem[{\citenamefont{Lewis et~al.}(1983)\citenamefont{Lewis, Davis,
  Archibald, Steinke, Smitherman, Doherty, Schnaper, LeWinter, Linares, Pouget
  et~al.}}]{Lewis1983}
\bibinfo{author}{\bibfnamefont{H.~D.} \bibnamefont{Lewis}},
  \bibinfo{author}{\bibfnamefont{J.~W.} \bibnamefont{Davis}},
  \bibinfo{author}{\bibfnamefont{D.~G.} \bibnamefont{Archibald}},
  \bibinfo{author}{\bibfnamefont{W.~E.} \bibnamefont{Steinke}},
  \bibinfo{author}{\bibfnamefont{T.~C.} \bibnamefont{Smitherman}},
  \bibinfo{author}{\bibfnamefont{J.~E.} \bibnamefont{Doherty}},
  \bibinfo{author}{\bibfnamefont{H.~W.} \bibnamefont{Schnaper}},
  \bibinfo{author}{\bibfnamefont{M.~M.} \bibnamefont{LeWinter}},
  \bibinfo{author}{\bibfnamefont{E.}~\bibnamefont{Linares}},
  \bibinfo{author}{\bibfnamefont{J.~M.} \bibnamefont{Pouget}},
  \bibnamefont{et~al.}, \bibinfo{journal}{N. Engl. J. Med.}
  \textbf{\bibinfo{volume}{309}}, \bibinfo{pages}{396} (\bibinfo{year}{1983}).

\bibitem[{\citenamefont{Chico et~al.}(1996)\citenamefont{Chico, Crespi,
  Benedict, Louie, and Cohen}}]{Chico1996}
\bibinfo{author}{\bibfnamefont{L.}~\bibnamefont{Chico}},
  \bibinfo{author}{\bibfnamefont{V.}~\bibnamefont{Crespi}},
  \bibinfo{author}{\bibfnamefont{L.}~\bibnamefont{Benedict}},
  \bibinfo{author}{\bibfnamefont{S.}~\bibnamefont{Louie}}, \bibnamefont{and}
  \bibinfo{author}{\bibfnamefont{M.}~\bibnamefont{Cohen}},
  \bibinfo{journal}{Phys. Rev. Lett.} \textbf{\bibinfo{volume}{76}},
  \bibinfo{pages}{971} (\bibinfo{year}{1996}).

\bibitem[{\citenamefont{Tans et~al.}(1998)\citenamefont{Tans, Verschueren, and
  Dekker}}]{Tans1998}
\bibinfo{author}{\bibfnamefont{S.~J.~S.} \bibnamefont{Tans}},
  \bibinfo{author}{\bibfnamefont{A.~R. M.~A.} \bibnamefont{Verschueren}},
  \bibnamefont{and} \bibinfo{author}{\bibfnamefont{C.}~\bibnamefont{Dekker}},
  \bibinfo{journal}{Nature} \textbf{\bibinfo{volume}{393}}, \bibinfo{pages}{49}
  (\bibinfo{year}{1998}).

\bibitem[{\citenamefont{Kwon et~al.}(1999)\citenamefont{Kwon, Tomanek, and
  Iijima}}]{Kwon1999}
\bibinfo{author}{\bibfnamefont{Y.-K.} \bibnamefont{Kwon}},
  \bibinfo{author}{\bibfnamefont{D.}~\bibnamefont{Tomanek}}, \bibnamefont{and}
  \bibinfo{author}{\bibfnamefont{S.}~\bibnamefont{Iijima}},
  \bibinfo{journal}{Phys. Rev. Lett.} \textbf{\bibinfo{volume}{82}},
  \bibinfo{pages}{1470} (\bibinfo{year}{1999}).

\bibitem[{\citenamefont{Z et~al.}(1999)\citenamefont{Z, Postma, L, and
  C}}]{Yao1999}
\bibinfo{author}{\bibfnamefont{Y.}~\bibnamefont{Z}},
  \bibinfo{author}{\bibfnamefont{H.}~\bibnamefont{Postma}},
  \bibinfo{author}{\bibfnamefont{B.}~\bibnamefont{L}}, \bibnamefont{and}
  \bibinfo{author}{\bibfnamefont{D.}~\bibnamefont{C}},
  \bibinfo{journal}{Nature} \textbf{\bibinfo{volume}{402}},
  \bibinfo{pages}{273} (\bibinfo{year}{1999}).

\bibitem[{\citenamefont{Sanvito et~al.}(1999)\citenamefont{Sanvito, Kwon,
  Tom\'{a}nek, and Lambert}}]{Sanvito1999}
\bibinfo{author}{\bibfnamefont{S.}~\bibnamefont{Sanvito}},
  \bibinfo{author}{\bibfnamefont{Y.-K.} \bibnamefont{Kwon}},
  \bibinfo{author}{\bibfnamefont{D.}~\bibnamefont{Tom\'{a}nek}},
  \bibnamefont{and} \bibinfo{author}{\bibfnamefont{C.~J.}
  \bibnamefont{Lambert}}, \bibinfo{journal}{Phys. Rev. Lett.}
  \textbf{\bibinfo{volume}{84}}, \bibinfo{pages}{1974} (\bibinfo{year}{1999}).

\bibitem[{\citenamefont{Berber et~al.}(2000)\citenamefont{Berber, Kwon, and
  Tomanek}}]{Berber2000}
\bibinfo{author}{\bibfnamefont{S.}~\bibnamefont{Berber}},
  \bibinfo{author}{\bibfnamefont{Y.-K.} \bibnamefont{Kwon}}, \bibnamefont{and}
  \bibinfo{author}{\bibfnamefont{D.}~\bibnamefont{Tomanek}},
  \bibinfo{journal}{Phys. Rev. Lett.} \textbf{\bibinfo{volume}{84}},
  \bibinfo{pages}{4613} (\bibinfo{year}{2000}).

\bibitem[{\citenamefont{Zhou et~al.}(2000)\citenamefont{Zhou, Kong, Yenilmez,
  and Dai}}]{Zhou2000}
\bibinfo{author}{\bibfnamefont{C.}~\bibnamefont{Zhou}},
  \bibinfo{author}{\bibfnamefont{J.}~\bibnamefont{Kong}},
  \bibinfo{author}{\bibfnamefont{E.}~\bibnamefont{Yenilmez}}, \bibnamefont{and}
  \bibinfo{author}{\bibfnamefont{H.}~\bibnamefont{Dai}},
  \bibinfo{journal}{Science (80-. ).} \textbf{\bibinfo{volume}{290}},
  \bibinfo{pages}{1552} (\bibinfo{year}{2000}).

\bibitem[{\citenamefont{Lee et~al.}(2002)\citenamefont{Lee, Kim, Kahng, Kim,
  Son, Ihm, Kato, Wang, Okazaki, Shinohara et~al.}}]{Lee2002}
\bibinfo{author}{\bibfnamefont{J.}~\bibnamefont{Lee}},
  \bibinfo{author}{\bibfnamefont{H.}~\bibnamefont{Kim}},
  \bibinfo{author}{\bibfnamefont{S.-J.} \bibnamefont{Kahng}},
  \bibinfo{author}{\bibfnamefont{G.}~\bibnamefont{Kim}},
  \bibinfo{author}{\bibfnamefont{Y.-W.} \bibnamefont{Son}},
  \bibinfo{author}{\bibfnamefont{J.}~\bibnamefont{Ihm}},
  \bibinfo{author}{\bibfnamefont{H.}~\bibnamefont{Kato}},
  \bibinfo{author}{\bibfnamefont{Z.~W.} \bibnamefont{Wang}},
  \bibinfo{author}{\bibfnamefont{T.}~\bibnamefont{Okazaki}},
  \bibinfo{author}{\bibfnamefont{H.}~\bibnamefont{Shinohara}},
  \bibnamefont{et~al.}, \bibinfo{journal}{Nature}
  \textbf{\bibinfo{volume}{415}}, \bibinfo{pages}{1005} (\bibinfo{year}{2002}).

\bibitem[{\citenamefont{Kim et~al.}(2005)\citenamefont{Kim, Lee, Kim, and
  Ihm}}]{Kim2005}
\bibinfo{author}{\bibfnamefont{G.}~\bibnamefont{Kim}},
  \bibinfo{author}{\bibfnamefont{S.~B.} \bibnamefont{Lee}},
  \bibinfo{author}{\bibfnamefont{T.~S.} \bibnamefont{Kim}}, \bibnamefont{and}
  \bibinfo{author}{\bibfnamefont{J.}~\bibnamefont{Ihm}},
  \bibinfo{journal}{Phys. Rev. B} \textbf{\bibinfo{volume}{71}}
  (\bibinfo{year}{2005}).

\bibitem[{\citenamefont{Charlier et~al.}(2007)\citenamefont{Charlier, Blase,
  and Roche}}]{Charlier2007}
\bibinfo{author}{\bibfnamefont{J.-C.} \bibnamefont{Charlier}},
  \bibinfo{author}{\bibfnamefont{X.}~\bibnamefont{Blase}}, \bibnamefont{and}
  \bibinfo{author}{\bibfnamefont{S.}~\bibnamefont{Roche}},
  \bibinfo{journal}{Rev. Mod. Phys.} \textbf{\bibinfo{volume}{79}},
  \bibinfo{pages}{677} (\bibinfo{year}{2007}).

\bibitem[{\citenamefont{Choi et~al.}(2008)\citenamefont{Choi, Ihm, and
  Kim}}]{Choi2008}
\bibinfo{author}{\bibfnamefont{W.~I.} \bibnamefont{Choi}},
  \bibinfo{author}{\bibfnamefont{J.}~\bibnamefont{Ihm}}, \bibnamefont{and}
  \bibinfo{author}{\bibfnamefont{G.}~\bibnamefont{Kim}},
  \bibinfo{journal}{Appl. Phys. Lett.} \textbf{\bibinfo{volume}{92}},
  \bibinfo{pages}{1} (\bibinfo{year}{2008}).

\bibitem[{\citenamefont{Long and Yang}(2001)}]{Long2001a}
\bibinfo{author}{\bibfnamefont{R.~Q.} \bibnamefont{Long}} \bibnamefont{and}
  \bibinfo{author}{\bibfnamefont{R.~T.} \bibnamefont{Yang}},
  \bibinfo{journal}{Ind. Eng. Chem. Res.} \textbf{\bibinfo{volume}{40}},
  \bibinfo{pages}{4288} (\bibinfo{year}{2001}).

\bibitem[{\citenamefont{Cinke et~al.}(2002)\citenamefont{Cinke, Li, Chen,
  Cassell, Delzeit, Han, and Meyyappan}}]{Cinke2002}
\bibinfo{author}{\bibfnamefont{M.}~\bibnamefont{Cinke}},
  \bibinfo{author}{\bibfnamefont{J.}~\bibnamefont{Li}},
  \bibinfo{author}{\bibfnamefont{B.}~\bibnamefont{Chen}},
  \bibinfo{author}{\bibfnamefont{A.}~\bibnamefont{Cassell}},
  \bibinfo{author}{\bibfnamefont{L.}~\bibnamefont{Delzeit}},
  \bibinfo{author}{\bibfnamefont{J.}~\bibnamefont{Han}}, \bibnamefont{and}
  \bibinfo{author}{\bibfnamefont{M.}~\bibnamefont{Meyyappan}},
  \bibinfo{journal}{Chem. Phys. Lett.} \textbf{\bibinfo{volume}{365}},
  \bibinfo{pages}{69} (\bibinfo{year}{2002}).

\bibitem[{\citenamefont{Yin et~al.}(2000)\citenamefont{Yin, Mays, and
  McEnaney}}]{Yin2000}
\bibinfo{author}{\bibfnamefont{Y.~F.} \bibnamefont{Yin}},
  \bibinfo{author}{\bibfnamefont{T.}~\bibnamefont{Mays}}, \bibnamefont{and}
  \bibinfo{author}{\bibfnamefont{B.}~\bibnamefont{McEnaney}},
  \bibinfo{journal}{Langmuir} \textbf{\bibinfo{volume}{16}},
  \bibinfo{pages}{10521} (\bibinfo{year}{2000}).

\bibitem[{\citenamefont{Dillon et~al.}(1997)\citenamefont{Dillon, Jones,
  Bekkedahl, Kiang, Bethune, and Heben}}]{Dillon1997}
\bibinfo{author}{\bibfnamefont{A.~C.} \bibnamefont{Dillon}},
  \bibinfo{author}{\bibfnamefont{K.~M.} \bibnamefont{Jones}},
  \bibinfo{author}{\bibfnamefont{T.~A.} \bibnamefont{Bekkedahl}},
  \bibinfo{author}{\bibfnamefont{C.~H.} \bibnamefont{Kiang}},
  \bibinfo{author}{\bibfnamefont{D.~S.} \bibnamefont{Bethune}},
  \bibnamefont{and} \bibinfo{author}{\bibfnamefont{M.~J.} \bibnamefont{Heben}},
  \bibinfo{journal}{Nature} \textbf{\bibinfo{volume}{386}},
  \bibinfo{pages}{377} (\bibinfo{year}{1997}).

\bibitem[{\citenamefont{Chambers et~al.}(1998)\citenamefont{Chambers, Park,
  Baker, and Rodriguez}}]{Chambers1998}
\bibinfo{author}{\bibfnamefont{A.}~\bibnamefont{Chambers}},
  \bibinfo{author}{\bibfnamefont{C.}~\bibnamefont{Park}},
  \bibinfo{author}{\bibfnamefont{R.~T.~K.} \bibnamefont{Baker}},
  \bibnamefont{and} \bibinfo{author}{\bibfnamefont{N.~M.}
  \bibnamefont{Rodriguez}}, \bibinfo{journal}{J. Phys. Chem. B}
  \textbf{\bibinfo{volume}{102}}, \bibinfo{pages}{4253} (\bibinfo{year}{1998}).

\bibitem[{\citenamefont{Peng et~al.}(2003)\citenamefont{Peng, Li, Luan, Di,
  Wang, Tian, and Jia}}]{Peng2003}
\bibinfo{author}{\bibfnamefont{X.}~\bibnamefont{Peng}},
  \bibinfo{author}{\bibfnamefont{Y.}~\bibnamefont{Li}},
  \bibinfo{author}{\bibfnamefont{Z.}~\bibnamefont{Luan}},
  \bibinfo{author}{\bibfnamefont{Z.}~\bibnamefont{Di}},
  \bibinfo{author}{\bibfnamefont{H.}~\bibnamefont{Wang}},
  \bibinfo{author}{\bibfnamefont{B.}~\bibnamefont{Tian}}, \bibnamefont{and}
  \bibinfo{author}{\bibfnamefont{Z.}~\bibnamefont{Jia}},
  \bibinfo{journal}{Chem. Phys. Lett.} \textbf{\bibinfo{volume}{376}},
  \bibinfo{pages}{154} (\bibinfo{year}{2003}).

\bibitem[{\citenamefont{Fagan et~al.}(2004)\citenamefont{Fagan, {Souza Filho},
  Lima, {Mendes Filho}, Ferreira, Mazali, Alves, and Dresselhaus}}]{Fagan2004}
\bibinfo{author}{\bibfnamefont{S.~B.} \bibnamefont{Fagan}},
  \bibinfo{author}{\bibfnamefont{A.~G.} \bibnamefont{{Souza Filho}}},
  \bibinfo{author}{\bibfnamefont{J.~O.~G.} \bibnamefont{Lima}},
  \bibinfo{author}{\bibfnamefont{J.}~\bibnamefont{{Mendes Filho}}},
  \bibinfo{author}{\bibfnamefont{O.~P.} \bibnamefont{Ferreira}},
  \bibinfo{author}{\bibfnamefont{I.~O.} \bibnamefont{Mazali}},
  \bibinfo{author}{\bibfnamefont{O.~L.} \bibnamefont{Alves}}, \bibnamefont{and}
  \bibinfo{author}{\bibfnamefont{M.~S.} \bibnamefont{Dresselhaus}},
  \bibinfo{journal}{Nano Lett.} \textbf{\bibinfo{volume}{4}},
  \bibinfo{pages}{1285} (\bibinfo{year}{2004}).

\bibitem[{\citenamefont{Fagan et~al.}(2006)\citenamefont{Fagan, Gir\~{a}o,
  Filho, and Filho}}]{Fagan2006}
\bibinfo{author}{\bibfnamefont{S.~B.} \bibnamefont{Fagan}},
  \bibinfo{author}{\bibfnamefont{E.~C.} \bibnamefont{Gir\~{a}o}},
  \bibinfo{author}{\bibfnamefont{J.~M.} \bibnamefont{Filho}}, \bibnamefont{and}
  \bibinfo{author}{\bibfnamefont{A.~G.~S.} \bibnamefont{Filho}},
  \bibinfo{journal}{Int. J. Quantum Chem.} \textbf{\bibinfo{volume}{106}},
  \bibinfo{pages}{2558} (\bibinfo{year}{2006}).

\bibitem[{\citenamefont{Tournus et~al.}(2005)\citenamefont{Tournus, Latil,
  Heggie, and Charlier}}]{Tournus2005}
\bibinfo{author}{\bibfnamefont{F.}~\bibnamefont{Tournus}},
  \bibinfo{author}{\bibfnamefont{S.}~\bibnamefont{Latil}},
  \bibinfo{author}{\bibfnamefont{M.}~\bibnamefont{Heggie}}, \bibnamefont{and}
  \bibinfo{author}{\bibfnamefont{J.-C.} \bibnamefont{Charlier}},
  \bibinfo{journal}{Phys. Rev. B} \textbf{\bibinfo{volume}{72}},
  \bibinfo{pages}{075431} (\bibinfo{year}{2005}).

\bibitem[{\citenamefont{Kong}(2000)}]{Kong2000}
\bibinfo{author}{\bibfnamefont{J.}~\bibnamefont{Kong}},
  \bibinfo{journal}{Science} \textbf{\bibinfo{volume}{287}},
  \bibinfo{pages}{622} (\bibinfo{year}{2000}).

\bibitem[{\citenamefont{Song et~al.}(2008)\citenamefont{Song, Lee, Jiang,
  Kussow, Lee, Hong, Kwon, and Choi}}]{Song2008}
\bibinfo{author}{\bibfnamefont{H.~J.} \bibnamefont{Song}},
  \bibinfo{author}{\bibfnamefont{Y.}~\bibnamefont{Lee}},
  \bibinfo{author}{\bibfnamefont{T.}~\bibnamefont{Jiang}},
  \bibinfo{author}{\bibfnamefont{A.~G.} \bibnamefont{Kussow}},
  \bibinfo{author}{\bibfnamefont{M.}~\bibnamefont{Lee}},
  \bibinfo{author}{\bibfnamefont{S.}~\bibnamefont{Hong}},
  \bibinfo{author}{\bibfnamefont{Y.-K.} \bibnamefont{Kwon}}, \bibnamefont{and}
  \bibinfo{author}{\bibfnamefont{H.~C.} \bibnamefont{Choi}},
  \bibinfo{journal}{J. Phys. Chem. C} \textbf{\bibinfo{volume}{112}},
  \bibinfo{pages}{629} (\bibinfo{year}{2008}).

\bibitem[{\citenamefont{Collins}(2000)}]{Collins2000}
\bibinfo{author}{\bibfnamefont{P.~G.} \bibnamefont{Collins}},
  \bibinfo{journal}{Science} \textbf{\bibinfo{volume}{287}},
  \bibinfo{pages}{1801} (\bibinfo{year}{2000}).

\bibitem[{\citenamefont{Bradley et~al.}(2003)\citenamefont{Bradley, Gabriel,
  Star, and Gr\"{u}ner}}]{Bradley2003}
\bibinfo{author}{\bibfnamefont{K.}~\bibnamefont{Bradley}},
  \bibinfo{author}{\bibfnamefont{J.~C.~P.} \bibnamefont{Gabriel}},
  \bibinfo{author}{\bibfnamefont{A.}~\bibnamefont{Star}}, \bibnamefont{and}
  \bibinfo{author}{\bibfnamefont{G.}~\bibnamefont{Gr\"{u}ner}},
  \bibinfo{journal}{Appl. Phys. Lett.} \textbf{\bibinfo{volume}{83}},
  \bibinfo{pages}{3821} (\bibinfo{year}{2003}).

\bibitem[{\citenamefont{Yu et~al.}(2000)\citenamefont{Yu, Files, Arepalli, and
  Ruoff}}]{Yu2000}
\bibinfo{author}{\bibfnamefont{M.-F.} \bibnamefont{Yu}},
  \bibinfo{author}{\bibfnamefont{B.~S.} \bibnamefont{Files}},
  \bibinfo{author}{\bibfnamefont{S.}~\bibnamefont{Arepalli}}, \bibnamefont{and}
  \bibinfo{author}{\bibfnamefont{R.~S.} \bibnamefont{Ruoff}},
  \bibinfo{journal}{Phys. Rev. Lett.} \textbf{\bibinfo{volume}{84}},
  \bibinfo{pages}{5552} (\bibinfo{year}{2000}).

\bibitem[{\citenamefont{Burt et~al.}(2005)\citenamefont{Burt, Wilson, Weaver,
  Dobson, and Macpherson}}]{Burt2005}
\bibinfo{author}{\bibfnamefont{D.~P.} \bibnamefont{Burt}},
  \bibinfo{author}{\bibfnamefont{N.~R.} \bibnamefont{Wilson}},
  \bibinfo{author}{\bibfnamefont{J.~M.~R.} \bibnamefont{Weaver}},
  \bibinfo{author}{\bibfnamefont{P.~S.} \bibnamefont{Dobson}},
  \bibnamefont{and} \bibinfo{author}{\bibfnamefont{J.~V.}
  \bibnamefont{Macpherson}}, \bibinfo{journal}{Nano Lett.}
  \textbf{\bibinfo{volume}{5}}, \bibinfo{pages}{639} (\bibinfo{year}{2005}).

\bibitem[{\citenamefont{Zhang et~al.}(2009)\citenamefont{Zhang, Thomas,
  Antonietti, and Wang}}]{Zhang2009}
\bibinfo{author}{\bibfnamefont{Y.}~\bibnamefont{Zhang}},
  \bibinfo{author}{\bibfnamefont{A.}~\bibnamefont{Thomas}},
  \bibinfo{author}{\bibfnamefont{M.}~\bibnamefont{Antonietti}},
  \bibnamefont{and} \bibinfo{author}{\bibfnamefont{X.}~\bibnamefont{Wang}},
  \bibinfo{journal}{J. Am. Chem. Soc.} \textbf{\bibinfo{volume}{131}},
  \bibinfo{pages}{50} (\bibinfo{year}{2009}).

\bibitem[{\citenamefont{Li et~al.}(1995)\citenamefont{Li, Chu, Cheng, Lin,
  Dravid, Chung, Wong, and Sproul}}]{Li1995}
\bibinfo{author}{\bibfnamefont{D.}~\bibnamefont{Li}},
  \bibinfo{author}{\bibfnamefont{X.}~\bibnamefont{Chu}},
  \bibinfo{author}{\bibfnamefont{S.-C.} \bibnamefont{Cheng}},
  \bibinfo{author}{\bibfnamefont{X.-W.} \bibnamefont{Lin}},
  \bibinfo{author}{\bibfnamefont{V.~P.} \bibnamefont{Dravid}},
  \bibinfo{author}{\bibfnamefont{Y.-W.} \bibnamefont{Chung}},
  \bibinfo{author}{\bibfnamefont{M.-S.} \bibnamefont{Wong}}, \bibnamefont{and}
  \bibinfo{author}{\bibfnamefont{W.~D.} \bibnamefont{Sproul}},
  \bibinfo{journal}{Appl. Phys. Lett.} \textbf{\bibinfo{volume}{67}},
  \bibinfo{pages}{203} (\bibinfo{year}{1995}).

\bibitem[{\citenamefont{Zambov et~al.}(2000)\citenamefont{Zambov, Popov,
  Abedinov, Plass, Kulisch, Gotszalk, Grabiec, Rangelow, and
  Kassing}}]{Zambov2000}
\bibinfo{author}{\bibfnamefont{B.~L.~M.} \bibnamefont{Zambov}},
  \bibinfo{author}{\bibfnamefont{C.}~\bibnamefont{Popov}},
  \bibinfo{author}{\bibfnamefont{N.}~\bibnamefont{Abedinov}},
  \bibinfo{author}{\bibfnamefont{M.~F.} \bibnamefont{Plass}},
  \bibinfo{author}{\bibfnamefont{W.}~\bibnamefont{Kulisch}},
  \bibinfo{author}{\bibfnamefont{T.}~\bibnamefont{Gotszalk}},
  \bibinfo{author}{\bibfnamefont{P.}~\bibnamefont{Grabiec}},
  \bibinfo{author}{\bibfnamefont{I.~W.} \bibnamefont{Rangelow}},
  \bibnamefont{and} \bibinfo{author}{\bibfnamefont{R.}~\bibnamefont{Kassing}},
  \bibinfo{journal}{Adv. Mater.} \textbf{\bibinfo{volume}{12}},
  \bibinfo{pages}{656} (\bibinfo{year}{2000}).

\bibitem[{\citenamefont{Cao et~al.}(2004)\citenamefont{Cao, Huang, Cao, Li, and
  Zhu}}]{Cao2004}
\bibinfo{author}{\bibfnamefont{C.}~\bibnamefont{Cao}},
  \bibinfo{author}{\bibfnamefont{F.}~\bibnamefont{Huang}},
  \bibinfo{author}{\bibfnamefont{C.}~\bibnamefont{Cao}},
  \bibinfo{author}{\bibfnamefont{J.}~\bibnamefont{Li}}, \bibnamefont{and}
  \bibinfo{author}{\bibfnamefont{H.}~\bibnamefont{Zhu}},
  \bibinfo{journal}{Chem. Mater.} \textbf{\bibinfo{volume}{16}},
  \bibinfo{pages}{5213} (\bibinfo{year}{2004}).

\bibitem[{\citenamefont{Guo et~al.}(2004)\citenamefont{Guo, Xie, Wang, Zhang,
  Hou, and Lv}}]{Guo2004}
\bibinfo{author}{\bibfnamefont{Q.}~\bibnamefont{Guo}},
  \bibinfo{author}{\bibfnamefont{Y.}~\bibnamefont{Xie}},
  \bibinfo{author}{\bibfnamefont{X.}~\bibnamefont{Wang}},
  \bibinfo{author}{\bibfnamefont{S.}~\bibnamefont{Zhang}},
  \bibinfo{author}{\bibfnamefont{T.}~\bibnamefont{Hou}}, \bibnamefont{and}
  \bibinfo{author}{\bibfnamefont{S.}~\bibnamefont{Lv}}, \bibinfo{journal}{Chem.
  Commun.} \textbf{\bibinfo{volume}{1}}, \bibinfo{pages}{26}
  (\bibinfo{year}{2004}).

\bibitem[{\citenamefont{Li et~al.}(2007)\citenamefont{Li, Cao, and
  Zhu}}]{Li2007}
\bibinfo{author}{\bibfnamefont{J.}~\bibnamefont{Li}},
  \bibinfo{author}{\bibfnamefont{C.}~\bibnamefont{Cao}}, \bibnamefont{and}
  \bibinfo{author}{\bibfnamefont{H.}~\bibnamefont{Zhu}},
  \bibinfo{journal}{Nanotechnology} \textbf{\bibinfo{volume}{18}},
  \bibinfo{pages}{115605} (\bibinfo{year}{2007}).

\bibitem[{\citenamefont{Pan et~al.}(2011)\citenamefont{Pan, Zhang, Shenoy, and
  Gao}}]{Pan2011}
\bibinfo{author}{\bibfnamefont{H.}~\bibnamefont{Pan}},
  \bibinfo{author}{\bibfnamefont{Y.-W.} \bibnamefont{Zhang}},
  \bibinfo{author}{\bibfnamefont{V.~B.} \bibnamefont{Shenoy}},
  \bibnamefont{and} \bibinfo{author}{\bibfnamefont{H.}~\bibnamefont{Gao}},
  \bibinfo{journal}{Nanoscale Res. Lett.} \textbf{\bibinfo{volume}{6}},
  \bibinfo{pages}{97} (\bibinfo{year}{2011}).

\bibitem[{\citenamefont{Hohenberg and Kohn}(1964)}]{Hohenberg1964}
\bibinfo{author}{\bibfnamefont{P.}~\bibnamefont{Hohenberg}} \bibnamefont{and}
  \bibinfo{author}{\bibfnamefont{W.}~\bibnamefont{Kohn}},
  \bibinfo{journal}{Phys. Rev. B} \textbf{\bibinfo{volume}{136}},
  \bibinfo{pages}{B864} (\bibinfo{year}{1964}).

\bibitem[{\citenamefont{Kohn and Sham}(1965)}]{Kohn1965}
\bibinfo{author}{\bibfnamefont{W.}~\bibnamefont{Kohn}} \bibnamefont{and}
  \bibinfo{author}{\bibfnamefont{L.~J.} \bibnamefont{Sham}},
  \bibinfo{journal}{Phys. Rev.} \textbf{\bibinfo{volume}{140}}
  (\bibinfo{year}{1965}).

\bibitem[{\citenamefont{Kresse and Hafner}(1993)}]{Kresse1993}
\bibinfo{author}{\bibfnamefont{G.}~\bibnamefont{Kresse}} \bibnamefont{and}
  \bibinfo{author}{\bibfnamefont{J.}~\bibnamefont{Hafner}},
  \bibinfo{journal}{Phys. Rev. B} \textbf{\bibinfo{volume}{47}},
  \bibinfo{pages}{558} (\bibinfo{year}{1993}).

\bibitem[{\citenamefont{Kresse}(1996)}]{Kresse1996}
\bibinfo{author}{\bibfnamefont{G.}~\bibnamefont{Kresse}},
  \bibinfo{journal}{Phys. Rev. B} \textbf{\bibinfo{volume}{54}},
  \bibinfo{pages}{11169} (\bibinfo{year}{1996}).

\bibitem[{\citenamefont{Perdew et~al.}(1996)\citenamefont{Perdew, Burke, and
  Ernzerhof}}]{Perdew1996}
\bibinfo{author}{\bibfnamefont{J.}~\bibnamefont{Perdew}},
  \bibinfo{author}{\bibfnamefont{K.}~\bibnamefont{Burke}}, \bibnamefont{and}
  \bibinfo{author}{\bibfnamefont{M.}~\bibnamefont{Ernzerhof}},
  \bibinfo{journal}{Phys. Rev. Lett.} \textbf{\bibinfo{volume}{77}},
  \bibinfo{pages}{3865} (\bibinfo{year}{1996}).

\bibitem[{\citenamefont{Grimme}(2006)}]{Grimme2006}
\bibinfo{author}{\bibfnamefont{S.}~\bibnamefont{Grimme}}, \bibinfo{journal}{J.
  Comput. Chem.} \textbf{\bibinfo{volume}{27}}, \bibinfo{pages}{1787}
  (\bibinfo{year}{2006}).

\bibitem[{\citenamefont{Monkhorst and Pack}(1976)}]{Monkhorst1976}
\bibinfo{author}{\bibfnamefont{H.~J.} \bibnamefont{Monkhorst}}
  \bibnamefont{and} \bibinfo{author}{\bibfnamefont{J.~D.} \bibnamefont{Pack}},
  \bibinfo{journal}{Phys. Rev. B} \textbf{\bibinfo{volume}{13}},
  \bibinfo{pages}{5188} (\bibinfo{year}{1976}).

\end{thebibliography}


@article{Artacho1999,
abstract = {A brief review of the Siesta project is presented in the context of linear-scaling density-functional methods for electronic-structure calculations and molecular-dynamics simulations of systems with a large number of atoms. Applications of the method to different systems are reviewed, including carbon nanotubes, gold nanostructures, adsorbates on silicon surfaces, and nucleic acids. Also, progress in atomic-orbital bases adapted to linear-scaling methodology is presented.},
author = {Artacho, E. and S\'{a}nchez-Portal, D. and Ordej\'{o}n, P. and Garc\'{\i}a, A. and Soler, J. M.},
doi = {10.1002/(SICI)1521-3951(199909)215:1<809::AID-PSSB809>3.0.CO;2-0},
isbn = {1521-3951},
issn = {1521-3951},
journal = {Phys. Status Solidi},
pages = {809--817},
title = {{Linear-Scaling ab-initio Calculations for Large and Complex Systems}},
volume = {215},
year = {1999}
}
@article{Berber2000,
abstract = {Combining equilibrium and nonequilibrium molecular dynamics simulations with accurate carbon potentials, we determine the thermal conductivity lambda of carbon nanotubes and its dependence on temperature. Our results suggest an unusually high value, lambda approximately 6600 W/m K, for an isolated (10,10) nanotube at room temperature, comparable to the thermal conductivity of a hypothetical isolated graphene monolayer or diamond. Our results suggest that these high values of lambda are associated with the large phonon mean free paths in these systems; substantially lower values are predicted and observed for the basal plane of bulk graphite.},
author = {Berber, S and Kwon, Y-K and Tomanek, D},
doi = {10.1103/PhysRevLett.84.4613},
journal = {Phys. Rev. Lett.},
pages = {4613--4616},
pmid = {10990753},
title = {{Unusually high thermal conductivity of carbon nanotubes}},
volume = {84},
year = {2000}
}
@article{Blase1994,
abstract = {Hybridization of the o* and x* states of the graphene network is shown to be as important as band-folding effects in determining the metallicity of small radius carbon nanotubes. Using detailed plane- wave ab initio pseudopotentia local density functional (LDA) calculations, we find that the electronic calculations. Strongly modified low-lying conduction band states are introduced into the band gap of insulating tubes because of strong a*-n* hybridization. As a result, the LDA gaps of some tubes are lowered by more than 50\%, and a tube previously predicted to be semiconducting is shown to be metallic.},
author = {Blase, X. and Benedict, Lorin X. and Shirley, Eric L. and Louie, Steven G.},
doi = {10.1103/PhysRevLett.72.1878},
journal = {Phys. Rev. Lett.},
number = {March},
pages = {1878--1881},
title = {{Hybridization effects and metallicity in small radius carbon nanotubes}},
volume = {72},
year = {1994}
}
@article{Bradley2003,
abstract = {We report a design for carbon nanotube field-effect transistors which tests the nanotube depletion length. In this design, the metal contacts and adjacent nanotubes were coated with impermeable silicon oxide, while the central region of nanotubes was exposed. We tested the devices by measuring sensitivity to NH3 and polyethylene imine.NH3 caused similar responses in passivated devices and in normal, nonpassivated devices. Thus, the device design passivates the metal-nanotube contacts while preserving chemical sensor characteristics. Polyethylene imine produced negative threshold shifts of tens of volts, despite being in contact with only the center region of devices. Based on the observed device characteristics, we conclude that the length scale of the covered nanotubes in our structure is comparable to the decay length of the depletion charge in nanotube transistors.},
author = {Bradley, Keith and Gabriel, Jean Christophe P and Star, Alexander and Gr\"{u}ner, George},
doi = {10.1063/1.1619222},
journal = {Appl. Phys. Lett.},
pages = {3821--3823},
title = {{Short-channel effects in contact-passivated nanotube chemical sensors}},
volume = {83},
year = {2003}
}
@article{Burt2005,
abstract = {We describe a method for the production of nanoelectrodes at the apex of atomic force microscopy (AFM) probes. The nanoelectrodes are formed from single-walled carbon nanotube AFM tips which act as the template for the formation of nanowire tips through sputter coating with metal. Subsequent deposition of a conformal insulating coating, and cutting of the probe end, yields a disk-shaped nanoelectrode at the AFM tip apex whose diameter is defined by the amount of metal deposited. We demonstrate that these probes are capable of high-resolution combined electrochemical and topographical imaging. The flexibility of this approach will allow the fabrication of nanoelectrodes of controllable size and composition, enabling the study of electrochemical activity at the nanoscale.},
author = {Burt, David P. and Wilson, Neil R. and Weaver, John M R and Dobson, Phillip S. and Macpherson, Julie V.},
doi = {10.1021/nl050018d},
journal = {Nano Lett.},
pages = {639--643},
title = {{Nanowire probes for high resolution combined scanning electrochemical microscopy - Atomic force microscopy}},
volume = {5},
year = {2005}
}
@article{Cao2004,
author = {Cao, Chuanbao and Huang, Fulin and Cao, Chuantang and Li, Jie and Zhu, Hesun},
doi = {10.1021/cm0493039},
journal = {Chem. Mater.},
number = {25},
pages = {5213--5215},
title = {{Synthesis of carbon nitride nanotubes via a catalytic-assembly solvothermal route}},
volume = {16},
year = {2004}
}
@article{Chambers1998,
abstract = {Graphite nanofibers are a novel material that is produced from the dissociation of carbon-containing gases over selected metal surfaces. The solid consists of very small graphite platelets, 30-500 Angstrom in width, which are stacked in a perfectly arranged conformation. We have discovered that the material is capable of sorbing and retaining in excess of 20 L (STP) of hydrogen per gram of carbon when the nanofibers are exposed to the gas at pressures of 120 atm at 25 degrees C, a value that is over an order of magnitude higher than that found with conventional hydrogen storage systems. This behavior is rationalized in terms of the unique crystalline arrangement existing within the graphite nanofiber structure, where the platelets generate a system comprised entirely of slit-shaped nanopores, in which only edge sites are exposed. Since the interplanar distance within the material is 3.37 Angstrom, sorption of molecular hydrogen, which possesses a kinetic diameter of 2.89 Angstrom, is a facile process owing to the short diffusion path. In addition, owing to the weak (van der Waals) bonding of the platelets, these nonrigid wall nanopores can expand to accommodate hydrogen in a multilayer configuration. Subsequent lowering of the pressure to nearly atmospheric conditions results in the release of a major fraction of the stored hydrogen at room temperature.},
author = {Chambers, Alan and Park, Colin and Baker, R Terry K and Rodriguez, Nelly M},
doi = {10.1021/jp980114l},
journal = {J. Phys. Chem. B},
pages = {4253--4256},
title = {{Hydrogen storage in graphite nanofibers}},
volume = {102},
year = {1998}
}
@article{Charlier2007,
abstract = {This article reviews the electronic and transport properties of carbon$\backslash$nnanotubes. The focus is mainly theoretical, but when appropriate$\backslash$nthe relation with experimental results is mentioned. While simple$\backslash$nband-folding arguments will be invoked to rationalize how the metallic$\backslash$nor semiconducting character of nanotubes is inferred from their topological$\backslash$nstructure, more sophisticated tight-binding and ab initio treatments$\backslash$nwill be introduced to discuss more subtle physical effects, such$\backslash$nas those induced by curvature, tube-tube interactions, or topological$\backslash$ndefects. The same approach will be followed for transport properties.$\backslash$nThe fundamental aspects of conduction regimes and transport length$\backslash$nscales will be presented using simple models of disorder, with the$\backslash$nderivation of a few analytic results concerning specific situations$\backslash$nof short- and long-range static perturbations. Further, the latest$\backslash$ndevelopments in semiempirical or ab initio simulations aimed at exploring$\backslash$nthe effect of realistic static scatterers (chemical impurities, adsorbed$\backslash$nmolecules, etc.) or inelastic electron-phonon interactions will be$\backslash$nemphasized. Finally, specific issues, going beyond the noninteracting$\backslash$nelectron model, will be addressed, including excitonic effects in$\backslash$noptical experiments, the Coulomb-blockade regime, and the Luttinger$\backslash$nliquid, charge density waves, or superconducting transition.},
author = {Charlier, J.-C. and Blase, X and Roche, S},
doi = {10.1103/RevModPhys.79.677},
journal = {Rev. Mod. Phys.},
keywords = {Electronic structure of nanoscale materials: clus,Nanotubes,and nanocrystals,nanoparticles,nanotubes},
pages = {677--732},
title = {{Electronic and transport properties of nanotubes}},
volume = {79},
year = {2007}
}
@article{Cheng2007,
author = {Cheng, Tsung O},
doi = {10.1111/j.1540-5834.2012.00707.x},
journal = {Texas Hear. Inst. J.},
number = {3},
pages = {392--393},
title = {{The history of aspirin.}},
volume = {34},
year = {2007}
}
@article{Chico1996,
abstract = {Introduction of pentagon-heptagon pair defects into the hexagonal network of a single carbon nanotube can change the helicity of the tube and alter its electronic structure. Using a tight-binding method to calculate the electronic structure of such systems we show that they behave as nanoscale metal/semiconductor or semiconductor/semiconductor junctions. These junctions could be the building blocks of nanoscale electronic devices made entirely of carbon.},
author = {Chico, L. and Crespi, Vincent and Benedict, Lorin and Louie, Steven and Cohen, Marvin},
doi = {10.1103/PhysRevLett.76.971},
journal = {Phys. Rev. Lett.},
pages = {971--974},
title = {{Pure Carbon Nanoscale Devices: Nanotube Heterojunctions}},
volume = {76},
year = {1996}
}
@article{Choi2008,
abstract = {We present the first-principles study of effects of the charge dopants such as Cesium and Iodine encapsulated on the electronic structure of carbon nanotubes. An encapsulated cesium atom donates an electron to the nanotube and produces donor-like states below the conduction bands. In contrast, an iodine trimer (I\$\_\{3\}\$) accepts an electron from the nanotube and produces an acceptor-like state above the valance band maximum. We find that a Cs atom inside a metallic armchair carbon nanotube gives rise to spatial oscillations of the density of states near the Fermi level.},
author = {Choi, Woon Ih and Ihm, Jisoon and Kim, Gunn},
doi = {10.1063/1.2929381},
journal = {Appl. Phys. Lett.},
number = {193110},
pages = {1--3},
title = {{Modification of the electronic structure in a carbon nanotube with the charge dopant encapsulation}},
volume = {92},
year = {2008}
}
@article{Cinke2002,
abstract = {Very high purity single-walled carbon nanotubes (SWNTs) were obtained from HiPco SWNT samples containing Fe particles by a two-step purification process. The raw and purified samples were characterized using high resolution transmission electron microscopy (HRTEM), Raman spectroscopy and thermogravimetric analysis (TGA). The purified sample consists of ???0.4\% Fe and the process does not seem to introduce any additional defects. The N 2 adsorption isotherm studies at 77 K reveal that the total surface area of the purified sample increases to 1587 m 2/g from 567 m 2/g for the raw material, which is the highest value reported for SWNTs. ?? 2002 Elsevier Science B.V. All rights reserved.},
author = {Cinke, Martin and Li, Jing and Chen, Bin and Cassell, Alan and Delzeit, Lance and Han, Jie and Meyyappan, M.},
doi = {10.1016/S0009-2614(02)01420-3},
journal = {Chem. Phys. Lett.},
pages = {69--74},
title = {{Pore structure of raw and purified HiPco single-walled carbon nanotubes}},
volume = {365},
year = {2002}
}
@article{Collins2000,
abstract = {The electronic properties of single-walled carbon nanotubes are shown here to be extremely sensitive to the chemical environment. Exposure to air or oxygen dramatically in\ss uences the nanotubes\~{O} electrical resistance, thermoelectric power, and local density of states, as determined by transport measurements and scanning tunneling spectroscopy. These electronic parameters can be reversibly \`{O}tuned\'{O} by surprisingly small concentrations of adsorbed gases, and an apparently semiconducting nanotube can be converted into an apparent metal through such exposure. These results, although demonstrating that nanotubes could Þnd use as sensitive chemical gas sensors, likewise indicate that many supposedly intrinsic properties measured on as-prepared nanotubes may be severely compromised by extrinsic air exposure effects.},
author = {Collins, P. G.},
doi = {10.1126/science.287.5459.1801},
journal = {Science},
pages = {1801--1804},
title = {{Extreme Oxygen Sensitivity of Electronic Properties of Carbon Nanotubes}},
volume = {287},
year = {2000}
}
@article{Dillon1997,
abstract = {Pores of molecular dimensions can adsorb large quantities of gases owing to the enhanced density of the adsorbed material inside the pores(1), a consequence of the attractive potential of the pore walls, Pederson and Broughton have suggested(2) that carbon nanotubes, which have diameters of typically a few nanometres, should be able to draw up liquids by capillarity, and this effect has been seen for low-surface-tension liquids in large-diameter, multi-walled nanotubes(3). Here we show that a gas can condense to high density inside narrow, single-walled nanotubes (SWNTs), Temperature-programmed desorption spectrosocopy shows that hydrogen will condense inside SWNTs under conditions that do not induce adsorption within a standard mesoporous activated carbon, The very high hydrogen uptake in these materials suggests that they might be effective as a hydrogen-storage material for fuel-cell electric vehicles},
author = {Dillon, A. C. and Jones, K. M. and Bekkedahl, T. A. and Kiang, C. H. and Bethune, D. S. and Heben, M. J.},
doi = {10.1038/386377a0},
journal = {Nature},
pages = {377--379},
title = {{Storage of hydrogen in single-walled carbon nanotubes}},
volume = {386},
year = {1997}
}
@article{Fagan2006,
abstract = {A systematic study of the structural and electronic properties of a 1,2- dichlorobenzene (DCB) molecule interacting with metallic single-wall carbon nanotubes is reported. The calculations were performed through ab initio methods using the SIESTA code. The interaction between DCB and nanotube is observed to depend on the diameter and it is larger for metallic nanotubes when compared with semiconducting. The binding energies are small thus suggesting that the interaction is through a physisorption process.},
author = {Fagan, Solange B. and Gir\~{a}o, E. C. and Filho, J. Mendes and Filho, A G Souza},
doi = {10.1002/qua.20962},
journal = {Int. J. Quantum Chem.},
keywords = {adsorption,carbon nanotubes,dichlorobenzene,electronic properties},
pages = {2558--2563},
title = {{First principles study of 1,2-dichlorobenzene adsorption on metallic carbon nanotubes}},
volume = {106},
year = {2006}
}
@article{Fagan2004,
abstract = {The interaction of 1,2-dichlorobenzene (DCB) with carbon nanotubes is analyzed by experimental and theoretical methods. Using first-principles calculations we studied the structural and electronic behavior of DCB interacting with a semiconductor (8,0) single-wall carbon nanotube (SWNT). We have found that the DCB weakly interacts with a perfect SWNT surface, but this interaction is slightly stronger when the SWNT surface has structural vacancies. Resonant Raman experiments performed on DCB-adsorbed SWNTs confirm the weak DCB-SWNT interaction, as suggested by the ab initio simulations.},
author = {Fagan, Solange B. and {Souza Filho}, A. G. and Lima, J. O G and {Mendes Filho}, J. and Ferreira, O. P. and Mazali, I. O. and Alves, O. L. and Dresselhaus, M. S.},
doi = {10.1021/nl0493895},
journal = {Nano Lett.},
pages = {1285--1288},
title = {{1,2-Dichlorobenzene interacting with carbon nanotubes}},
volume = {4},
year = {2004}
}
@article{Ferreira1973,
abstract = {1. Resting splenic venous outflow from anaesthetized dogs contains prostaglandin-like material: the concentration increases after intra-arterial injections of bradykinin into the spleen, and is abolished by treatment with indomethacin. 2. Intra-arterial injections of bradykinin into the spleen of lightly anaesthetized dogs elicit a dose-dependent reflex increase in the blood pressure, which is reduced but not abolished by treatment with indomethacin. 3. Addition of prostaglandin E1 or E2 either by injections or by infusions restores the reflex increase in the blood pressure due to bradykinin injections after indomethacin treatment. 4. The sensitizing action of endogenously released prostaglandins at or near the afferent nerve endings is discussed. 5. The analgesic activity of aspirin-like drugs is explained in terms of the removal of the sensitizing activity of prostaglandins.},
author = {Ferreira, S. H. and Moncada, S. and Vane, J. R.},
doi = {10.1111/j.1476-5381.1997.tb06822.x},
journal = {Br. J. Pharmacol.},
pages = {86--97},
title = {{Prostaglandins and the mechanism of analgesis produced by aspirin-like drugs}},
volume = {49},
year = {1973}
}
@article{Gracia2009,
abstract = {Gracia, J., \& Kroll, P. (2009). First principles study of C3N4 carbon nitride nanotubes. Journal of Materials Chemistry, 19, 3020. doi:10.1039/b821569cWe investigate the structural and optical properties of carbon nitride (C3N4) nanotubes (CNNts) built from condensed heptazine cores (C6N7) with different chirality and connection patterns. In particular, heterocycles in a hexagonal ordering exhibit the lowest energy configuration for the CNNts studied. Overall, heptazine-based CNNts are energetically preferred over triazine motifs. Correspondingly, recent experimental data show the prevalence of heptazine units in synthesized CNNts. Nitrogen-nitrogen lone pair repulsions prevent CNNts from presenting smooth tubular surfaces. Corrugation in general stabilizes C3N4 nano-structures with respect to extended conformations in comparison with pure carbon compositions. In connection to experiment, hexagonal nanotubes show optical properties that are almost independent of the chirality and tube diameter. CNNts show potential for similar applications as carbon nanotubes, and may even improve on the performance in some specific areas, as they have stable semiconducting parameters, and have polarized C–N bonds together with surface holes, which suggest them as better hosts as storage devices.},
author = {Gracia, Jose and Kroll, Peter},
journal = {J. Mater. Chem.},
pages = {3020},
title = {{First principles study of C3N4 carbon nitride nanotubes}},
volume = {19},
year = {2009}
}
@article{Grimme2006,
abstract = {A new density functional (DF) of the generalized gradient approximation (GGA) type for general chemistry applications termed B97-D is proposed. It is based on Becke's power-series ansatz from 1997 and is explicitly parameterized by including damped atom-pairwise dispersion corrections of the form C(6) x R(-6). A general computational scheme for the parameters used in this correction has been established and parameters for elements up to xenon and a scaling factor for the dispersion part for several common density functionals (BLYP, PBE, TPSS, B3LYP) are reported. The new functional is tested in comparison with other GGAs and the B3LYP hybrid functional on standard thermochemical benchmark sets, for 40 noncovalently bound complexes, including large stacked aromatic molecules and group II element clusters, and for the computation of molecular geometries. Further cross-validation tests were performed for organometallic reactions and other difficult problems for standard functionals. In summary, it is found that B97-D belongs to one of the most accurate general purpose GGAs, reaching, for example for the G97/2 set of heat of formations, a mean absolute deviation of only 3.8 kcal mol(-1). The performance for noncovalently bound systems including many pure van der Waals complexes is exceptionally good, reaching on the average CCSD(T) accuracy. The basic strategy in the development to restrict the density functional description to shorter electron correlation lengths scales and to describe situations with medium to large interatomic distances by damped C(6) x R(-6) terms seems to be very successful, as demonstrated for some notoriously difficult reactions. As an example, for the isomerization of larger branched to linear alkanes, B97-D is the only DF available that yields the right sign for the energy difference. From a practical point of view, the new functional seems to be quite robust and it is thus suggested as an efficient and accurate quantum chemical method for large systems where dispersion forces are of general importance.},
author = {Grimme, Stefan},
doi = {10.1002/jcc.20495},
journal = {J. Comput. Chem.},
keywords = {Density functional theory,Generalized gradient approximation,Thermochemistry,Van der Waals interactions},
pages = {1787--1799},
title = {{Semiempirical GGA-type density functional constructed with a long-range dispersion correction}},
volume = {27},
year = {2006}
}
@article{Gulseren2002,
abstract = {We investigate curvature effects on geometric parameters, energetics, and electronic structure of zigzag nanotubes with fully optimized geometries from first-principle calculations. The calculated curvature energies, which are inversely proportional to the square of radius, are in good agreement with the classical elasticity theory. The variation of the band gap with radius is found to differ from simple rules based on the zone folded graphene bands. Large discrepancies between tight binding and first-principles calculations of the band gap values of small nanotubes are discussed in detail.},
author = {G\"{u}lseren, O. and Yildirim, T. and Ciraci, S.},
doi = {10.1103/PhysRevB.65.153405},
journal = {Phys. Rev. B},
pages = {153405},
title = {{Systematic ab initio study of curvature effects in carbon nanotubes}},
volume = {65},
year = {2002}
}
@article{Guo2004,
author = {Guo, Qixun and Xie, Yi and Wang, Xinjun and Zhang, Shuyuan and Hou, Tao and Lv, Shichang},
doi = {10.1039/C4RA13342K},
journal = {Chem. Commun.},
number = {2},
pages = {26--27},
title = {{Synthesis of carbon nitride nanotubes with the C3N4 stoichiometry via a benzene-thermal process at low temperatures}},
volume = {1},
year = {2004}
}
@article{Hernandez1998,
abstract = {We present a comparative study of the energetic, structural, and elastic properties of carbon and composite single-wall nanotubes, including BN, BC3, and BC2N nanotubes, using a nonorthogonal tight-binding formalism. Our calculations predict that carbon nanotubes have a higher Young modulus than any of the studied composite nanotubes, and of the same order as that found for defect-free graphene sheets. We obtain good agreement with the available experimental results.},
author = {Hern\'{a}ndez, E. and Goze, C. and Bernier, P. and Rubio, A.},
doi = {10.1103/PhysRevLett.80.4502},
journal = {Phys. Rev. Lett.},
pages = {4502--4505},
title = {{Elastic properties of C and \$B\_xC\_yN\_z\$ composite nanotubes}},
volume = {80},
year = {1998}
}
@article{Hohenberg1964,
abstract = {This paper deals with the ground state of an interacting electron gas in an external potential v(r). It is proved that there exists a universal functional of the density, F[n(r)], independent of v(r), such that the expression E≡∫v(r)n(r)dr+F[n(r)] has as its minimum value the correct ground-state energy associated with v(r). The functional F[n(r)] is then discussed for two situations: (1) n(r)=n0+ñ(r), ñ/n0≪1, and (2) n(r)=ϕ(r/r0) with ϕ arbitrary and r0→∞. In both cases F can be expressed entirely in terms of the correlation energy and linear and higher order electronic polarizabilities of a uniform electron gas. This approach also sheds some light on generalized Thomas-Fermi methods and their limitations. Some new extensions of these methods are presented.},
author = {Hohenberg, P. and Kohn, W.},
journal = {Phys. Rev. B},
number = {3B},
pages = {B864--B871},
title = {{Inhomogeneous electron gas}},
volume = {136},
year = {1964}
}
@misc{Iijima1991,
abstract = {The synthesis of molecular carbon structures in the form of C60 and other fullerenes1 has stimulated intense interest in the structures accessible to graphitic carbon sheets. Here I report the preparation of a new type of finite carbon structure consisting of needle-like tubes. Produced using an are-discharge evaporation method similar to that used for fullerene synthesis, the needles grow at the negative end of the electrode used for the are discharge. Electron microscopy reveals that each needle comprises coaxial tubes of graphitic sheets, ranging in number from 2 up to about 50. On each tube the carbon-atom hexagons are arranged in a helical fashion about the needle axis. The helical pitch varies from needle to needle and from tube to tube within a single needle. It appears that this helical structure may aid the growth process. The formation of these needles, ranging from a few to a few tens of nanometres in diameter, suggests that engineering of carbon structures should be possible on scales considerably greater than those relevant to the fullerenes.},
author = {Iijima, Sumo},
booktitle = {Nature},
doi = {10.1038/354056a0},
pages = {56--58},
title = {{Helical microtubules of graphitic carbon}},
volume = {354},
year = {1991}
}
@article{Ito2002,
author = {Ito, T and Nishidate, K and Baba, M and Hasegawa, M},
journal = {Surf. Sci.},
keywords = {carbon,computer simulations,density functional calculations,surface energy,surface stress},
pages = {222--226},
title = {{First principles calculations for electronic band structure of single-walled carbon nanotube under uniaxial strain}},
volume = {V514},
year = {2002}
}
@article{Jin2003,
author = {Jin, Y},
doi = {10.1016/S0266-3538(03)00074-5},
journal = {Compos. Sci. Technol.},
pages = {1507--1515},
title = {{Simulation of elastic properties of single-walled carbon nanotubes}},
volume = {63},
year = {2003}
}
@article{Kim2005,
abstract = {We investigate the electron transport in multiply connected metallic carbon nanotubes within the Landauer-B$\backslash$"\{u\}ttiker formalism. Quasibound states coupled to the incident \$\backslash pi\^{}\{*\}\$ states give rise to energy levels of different widths depending on the coupling strength. In particular, donorlike states originating from heptagonal rings are found to give a very narrow level. Interference between broad and narrow levels produces Fano-type resonant backscattering as well as resonant tunneling. Over a significantly wide energy range, almost perfect suppression of the conduction of \$\backslash pi\^{}\{*\}\$ electrons occurs, which may be regarded as filtering of particular electrons (\$\backslash pi\$-pass filter).},
author = {Kim, Gunn and Lee, Sang Bong and Kim, Tae Suk and Ihm, Jisoon},
doi = {10.1103/PhysRevB.71.205415},
journal = {Phys. Rev. B},
title = {{Fano resonance and orbital filtering in multiply connected carbon nanotubes}},
volume = {71},
year = {2005}
}
@article{Kohn1965,
abstract = {From a theory of Hohenberg and Kohn, approximation methods for treating an inhomogeneous system of interacting electrons are developed. These methods are exact for systems of slowly varying or high density. For the ground state, they lead to self-consistent equations analogous to the Hartree and \{Hartree-Fock\} equations, respectively. In these equations the exchange and correlation portions of the chemical potential of a uniform electron gas appear as additional effective potentials. \{(The\} exchange portion of our effective potential differs from that due to Slater by a factor of 2 / 3.) Electronic systems at finite temperatures and in magnetic fields are also treated by similar methods. An appendix deals with a further correction for systems with short-wavelength density oscillations.},
author = {Kohn, W. and Sham, L. J.},
doi = {10.1103/PhysRev.140.A1133},
journal = {Phys. Rev.},
title = {{Self-consistent equations including exchange and correlation effects}},
volume = {140},
year = {1965}
}
@article{Kong2000,
abstract = {Chemical sensors based on individual single-walled carbon nanotubes (SWNTs) are demonstrated. Upon exposure to gaseous molecules such as NO(2) or NH(3), the electrical resistance of a semiconducting SWNT is found to dramatically increase or decrease. This serves as the basis for nanotube molecular sensors. The nanotube sensors exhibit a fast response and a substantially higher sensitivity than that of existing solid-state sensors at room temperature. Sensor reversibility is achieved by slow recovery under ambient conditions or by heating to high temperatures. The interactions between molecular species and SWNTs and the mechanisms of molecular sensing with nanotube molecular wires are investigated.},
author = {Kong, J.},
doi = {10.1126/science.287.5453.622},
journal = {Science},
pages = {622--625},
title = {{Nanotube Molecular Wires as Chemical Sensors}},
volume = {287},
year = {2000}
}
@article{Kozinsky2006,
abstract = {We characterize the response of isolated single-wall (SWNT) and multiwall (MWNT) carbon nanotubes and nanotube bundles to static electric fields using first-principles calculations and density-functional theory. The longitudinal polarizability of SWNTs scales as the inverse square of the band gap, while in MWNTs and bundles it is given by the sum of the polarizabilities of the constituent tubes. The transverse polarizability of SWNTs is insensitive to band gaps and chiralities and is proportional to the square of the effective radius; in MWNTs, the outer layers dominate the response. The transverse response is intermediate between metallic and insulating, and a simple electrostatic model based on a scale-invariance relation captures accurately the first-principles results. The dielectric response of nonchiral SWNTs in both directions remains linear up to very high values of applied field.},
author = {Kozinsky, Boris and Marzari, Nicola},
doi = {10.1103/PhysRevLett.96.166801},
journal = {Phys. Rev. Lett.},
number = {April},
pages = {2--5},
title = {{Static dielectric properties of carbon nanotubes from first principles}},
volume = {96},
year = {2006}
}
@article{Kresse1996,
abstract = {We present an efficient scheme for calculating the Kohn-Sham ground state of metallic systems using pseudopotentials and a plane-wave basis set. In the first part the application of Pulay’s DIIS method (direct inversion in the iterative subspace) to the iterative diagonalization of large matrices will be discussed. Our approach is stable, reliable, and minimizes the number of order Natoms3 operations. In the second part, we will discuss an efficient mixing scheme also based on Pulay’s scheme. A special ‘‘metric’’ and a special ‘‘preconditioning’’ optimized for a plane-wave basis set will be introduced. Scaling of the method will be discussed in detail for non-self-consistent and self-consistent calculations. It will be shown that the number of iterations required to obtain a specific precision is almost independent of the system size. Altogether an order Natoms2 scaling is found for systems containing up to 1000 electrons. If we take into account that the number of k points can be decreased linearly with the system size, the overall scaling can approach Natoms. We have implemented these algorithms within a powerful package called VASP (Vienna ab initio simulation package). The program and the techniques have been used successfully for a large number of different systems (liquid and amorphous semiconductors, liquid simple and transition metals, metallic and semiconducting surfaces, phonons in simple metals, transition metals, and semiconductors) and turned out to be very reliable. © 1996 The American Physical Society.},
author = {Kresse, G.},
doi = {10.1103/PhysRevB.54.11169},
journal = {Phys. Rev. B},
pages = {11169--11186},
title = {{Efficient iterative schemes for ab initio total-energy calculations using a plane-wave basis set}},
volume = {54},
year = {1996}
}
@article{Kresse1993,
abstract = {We present ab initio quantum-mechanical molecular-dynamics calculations based on the calculation of the electronic ground state and of the Hellmann-Feynman forces in the local-density approximation at each molecular-dynamics step. This is possible using conjugate-gradient techniques for energy minimization, and predicting the wave functions for new ionic positions using sub-space lignment. This approach avoids the instabilities inherent in quantum-mechanical molecular-dynamics calculations for metals based on the use of a fictitious Newtonian dynamics for the electronic degree of freedom. This method gives perfect control of the adiabaticity and allows us to perform simulations over several picoseconds.},
author = {Kresse, G. and Hafner, J.},
doi = {10.1103/PhysRevB.47.558},
journal = {Phys. Rev. B},
pages = {558--561},
title = {{Ab initio molecular dynamics for liquid metals}},
volume = {47},
year = {1993}
}
@article{Krishnan1998,
abstract = {We estimate the stiffness of single-walled carbon nanotubes by observing their freestanding room- temperature vibrations in a transmission electron microscope. The nanotube dimensions and vibration ampli- tude are measured from electron micrographs, and it is assumed that the vibration modes are driven stochas- tically and are those of a clamped cantilever. Micrographs of 27 nanotubes in the diameter range 1.0–1.5 nm were measured to yield an average Young’s modulus of Y1.25 TPa. This value is consistent with previous measurements for multiwalled nanotubes, and is higher than the currently accepted value of the in-plane modulus of graphite.},
author = {Krishnan, a. and Dujardin, E. and Ebbesen, T. and Yianilos, P. and Treacy, M.},
doi = {10.1103/PhysRevB.58.14013},
file = {:C$\backslash$:/Users/yjlee/Desktop/CNT논문 Ref/youngref/PhysRevB.58.14013.pdf:pdf},
journal = {Phys. Rev. B},
number = {20},
pages = {14013--14019},
title = {{Young’s modulus of single-walled nanotubes}},
volume = {58},
year = {1998}
}
@article{Kwon1999,
abstract = {Thermal treatment is reported to convert finely dispersed diamond powder to multiwall carbon nanocapsules containing fullerenes such as C-60. We investigate the internal dynamics of a related model system, consisting of a K@C-60(+) endohedral complex enclosed in a C-480 nanocapsule. We show this to be a tunable two-level system, where transitions between the two states can be induced by applying an electric field between the C-480 end caps, and discuss its potential application as a nonvolatile memory element. [S0031-9007(99)08484-7].},
author = {Kwon, Y-K and Tomanek, D and Iijima, S},
journal = {Phys. Rev. Lett.},
pages = {1470--1473},
title = {{"Bucky shuttle" memory device: Synthetic approach and molecular dynamics simulations}},
volume = {82},
year = {1999}
}
@article{Lee2002,
abstract = {Motivated by the technical and economic difficulties in further miniaturizing silicon-based transistors with the present fabrication technologies, there is a strong effort to develop alternative electronic devices, based, for example, on single molecules. Recently, carbon nanotubes have been successfully used for nanometre-sized devices such as diodes, transistors, and random access memory cells. Such nanotube devices are usually very long compared to silicon-based transistors. Here we report a method for dividing a semiconductor nanotube into multiple quantum dots with lengths of about 10nm by inserting Gd@C82 endohedral fullerenes. The spatial modulation of the nanotube electronic bandgap is observed with a low-temperature scanning tunnelling microscope. We find that a bandgap of approximately 0.5eV is narrowed down to approximately 0.1eV at sites where endohedral metallofullerenes are inserted. This change in bandgap can be explained by local elastic strain and charge transfer at metallofullerene sites. This technique for fabricating an array of quantum dots could be used for nano-electronics and nano-optoelectronics.},
author = {Lee, Jhinhwan and Kim, H and Kahng, S-J and Kim, G and Son, Y-W and Ihm, J and Kato, H and Wang, Z W and Okazaki, T and Shinohara, H and Kuk, Young},
doi = {10.1038/4151005a},
journal = {Nature},
pages = {1005--1008},
title = {{Bandgap modulation of carbon nanotubes by encapsulated metallofullerenes.}},
volume = {415},
year = {2002}
}
@article{Lewis1983,
abstract = {We conducted a multicenter, double-blind, placebo-controlled randomized trial of aspirin treatment (324 mg in buffered solution daily) for 12 weeks in 1266 men with unstable angina (625 taking aspirin and 641 placebo). The principal end points were death and acute myocardial infarction diagnosed by the presence of creatine kinase MB or pathologic Q-wave changes on electrocardiograms. The incidence of death or acute myocardial infarction was 51 per cent lower in the aspirin group than in the placebo group: 31 patients (5.0 per cent) as compared with 65 (10.1 per cent); P = 0.0005. Nonfatal acute myocardial infarction was 51 per cent lower in the aspirin group: 21 patients (3.4 per cent) as compared with 44 (6.9 per cent); P = 0.005. The reduction in mortality in the aspirin group was also 51 per cent--10 patients (1.6 per cent) as compared with 21 (3.3 per cent)--although it was not statistically significant; P = 0.054. There was no difference in gastrointestinal symptoms or evidence of blood loss between the treatment and control groups. Our data show that aspirin has a protective effect against acute myocardial infarction in men with unstable angina, and they suggest a similar effect on mortality.},
author = {Lewis, H D and Davis, J W and Archibald, D G and Steinke, W E and Smitherman, T C and Doherty, J E and Schnaper, H W and LeWinter, M M and Linares, E and Pouget, J M and Sabharwal, S C and Chesler, E and DeMots, H},
doi = {10.1056/NEJM198308183090703},
journal = {N. Engl. J. Med.},
pages = {396--403},
title = {{Protective effects of aspirin against acute myocardial infarction and death in men with unstable angina. Results of a Veterans Administration Cooperative Study.}},
volume = {309},
year = {1983}
}
@article{Li2007,
author = {Li, Jie and Cao, Chuanbao and Zhu, Hesun},
doi = {10.1088/0957-4484/18/11/115605},
journal = {Nanotechnology},
pages = {115605},
title = {{Synthesis and characterization of graphite-like carbon nitride nanobelts and nanotubes}},
volume = {18},
year = {2007}
}
@article{Long2001a,
abstract = {Carbon nanotubes were prepared and investigated as sorbents for NO, SO2, and CO2 in the presence of O-2. The preliminary results indicate that they are a good and reversible sorbent for NO removal at room temperature. An uptake amount of 78 mg/g of NOx was obtained by TGA when the carbon nanotubes were exposed to 1000 ppm NO + 5\% O-2/He for 120 min. The equilibrium amount was near 90 mg/g after 12 h. TPD profiles show that NO2 and NO desorb from the NOx-saturated carbon nanotubes at well below 300 degreesC. FTIR spectra indicate that the adsorbed species are nitrates and trace amounts of NO2 and (NO)(2) dimers. By comparison, only 29 mg/g of SO2 (at 500 ppm) and 2 mg/g of CO2 (at 10\%) are adsorbed onto the carbon nanotubes in 120 min.},
author = {Long, R Q and Yang, R T},
journal = {Ind. Eng. Chem. Res.},
pages = {4288--4291},
title = {{Carbon nanotubes as a superior sorbent for nitrogen oxides}},
volume = {40},
year = {2001}
}
@article{Long2001a,
author = {Long, R Q and Yang, R T},
journal = {Jounal Am. Chem. Soc.},
number = {9},
pages = {2058--2059},
title = {{Carbon Nanotubes as Superior Sorbent for Dioxin Removal}},
volume = {123},
year = {2001}
}
@article{Lu1997,
abstract = {Elastic properties of carbon nanotubes and nanoropes are investigated using an empirical force-constant model. For single and multiwall nanotubes the elastic moduli are shown to be insensitive to structural details such as the helicity, the radius, and the number of walls. The tensile Young's modulus and the torsion shear modulus of tubes are comparable to that of the diamond, while the bulk modulus is smaller. Nanoropes composed of single wall nanotubes have the ideal elastic properties of high tensile stiffness and light weight.},
author = {Lu, Jian Ping},
doi = {10.1103/PhysRevLett.79.1297},
journal = {Phys. Rev. Lett.},
pages = {1297--1300},
title = {{Elastic properties of carbon nanotubes and nanoropes}},
volume = {79},
year = {1997}
}
@article{Mintmire1992,
abstract = {We have calculated the electronic structure of a fullerene tubule using a first-principles, self-consistent, all-electron Gaussian-orbital based local-density-functional approach. Extending these results to a model containing an electron-lattice interaction, we estimate that the mean-field transition temperature from a Peierls-distorted regime to a high-temperature metallic regime should be well below room temperature. Such fullerene tubules should have the advantages (compared to the other conjugated carbon systems) of a carrier density similar to that of metals and zero band gap at room temperature.},
author = {Mintmire, J. W. and Dunlap, B. I. and White, C. T.},
doi = {10.1103/PhysRevLett.68.631},
journal = {Phys. Rev. Lett.},
number = {5},
pages = {631--634},
title = {{Are fullerene tubules metallic?}},
volume = {68},
year = {1992}
}
@article{Monkhorst1976,
author = {Monkhorst, Hendrik J. and Pack, James D.},
journal = {Phys. Rev. B},
pages = {5188--5192},
title = {{Special points for brillouin-zone integrations}},
volume = {13},
year = {1976}
}
@article{Odom1998,
abstract = {Carbon nanotubes1 are predicted to be metallic or semiconducting depending on their diameter and the helicity of the arrangement of graphitic rings in their walls2, 3, 4, 5. Scanning tunnelling microscopy (STM) offers the potential to probe this prediction, as it can resolve simultaneously both atomic structure and the electronic density of states. Previous STM studies of multi-walled nanotubes6, 7, 8, 9 and single-walled nanotubes (SWNTs)10 have provided indications of differing structures and diameter-dependent electronic properties, but have not revealed any explicit relationship between structure and electronic properties. Here we report STM measurements of the atomic structure and electronic properties of SWNTs. We are able to resolve the hexagonal-ring structure of the walls, and show that the electronic properties do indeed depend on diameter and helicity. We find that the SWNT samples exhibit many different structures, with no one species dominating.},
author = {Odom, T.W. and Huang, J.L. and Kim, P. and Lieber, C.M.},
doi = {10.1038/34139},
journal = {Nature},
pages = {62--64},
title = {{Atomic structure and electronic properties of single-walled carbon nanotubes}},
volume = {391},
year = {1998}
}
@article{Pan2011,
abstract = {ABSTRACT: The magnetic properties of metal-functionalized graphitic carbon nitride nanotubes were investigated based on first-principles calculations. The graphitic carbon nitride nanotube can be either ferromagnetic or antiferromagnetic by functionalizing with different metal atoms. The W- and Ti-functionalized nanotubes are ferromagnetic, which are attributed to carrier-mediated interactions because of the coupling between the spin-polarized d and p electrons and the formation of the impurity bands close to the band edges. However, Cr-, Mn-, Co-, and Ni-functionalized nanotubes are antiferromagnetic because of the anti-alignment of the magnetic moments between neighboring metal atoms. The functionalized nanotubes may be used in spintronics and hydrogen storage.},
author = {Pan, Hui and Zhang, Yong-Wei and Shenoy, Vivek B and Gao, Huajian},
doi = {10.1186/1556-276X-6-97},
journal = {Nanoscale Res. Lett.},
number = {1},
pages = {97},
publisher = {Springer Open Ltd},
title = {{Metal-functionalized single-walled graphitic carbon nitride nanotubes: a first-principles study on magnetic property.}},
volume = {6},
year = {2011}
}
@article{Peng2003,
abstract = {The as-grown CNTs and graphitized CNTs were used as adsorbents to remove 1,2-dichlorobenzene from water. The experiments demonstrate that it takes only 40 min for CNTs to attain equilibrium and the adsorption capacity of as-grown and graphitized CNTs is 30.8 and 28.7 mg/g, respectively, from a 20 mg/l solution. CNTs can be used as adsorbents in a wide pH range of 3-10. Thermodynamic calculations indicate that the adsorption reaction is spontaneous with a high affinity and the adsorption is an endothermic reaction. ?? 2003 Elsevier Science B.V. All rights reserved.},
author = {Peng, Xianjia and Li, Yanhui and Luan, Zhaokun and Di, Zechao and Wang, Hongyu and Tian, Binghui and Jia, Zhiping},
doi = {10.1016/S0009-2614(03)00960-6},
journal = {Chem. Phys. Lett.},
pages = {154--158},
title = {{Adsorption of 1,2-dichlorobenzene from water to carbon nanotubes}},
volume = {376},
year = {2003}
}
@article{Perdew1996,
abstract = {Generalized gradient approximations (GGA's) for the exchange-correlation energy improve upon the local spin density (LSD) description of atoms, molecules, and solids. We present a simple derivation of a simple GGA, in which all parameters (other than those in LSD) are fundamental constants. Only general features of the detailed construction underlying the Perdew-Wang 1991 (PW91) GGA are invoked. Improvements over PW91 include an accurate description of the linear response of the uniform electron gas, correct behavior under uniform scaling, and a smoother potential.},
author = {Perdew, Jp and Burke, K and Ernzerhof, M},
doi = {10.1103/PhysRevLett.77.3865},
journal = {Phys. Rev. Lett.},
pages = {3865--3868},
title = {{Generalized Gradient Approximation Made Simple.}},
volume = {77},
year = {1996}
}
@article{Sanchez-Portal1997,
abstract = {We have implemented a linear scaling, fully self-consistent density-functional method for performing first-principles calculations on systems with a large number of atoms, using standard norm-conserving pseudopotentials and flexible linear combinations of atomic orbitals (LCAO) basis sets. Exchange and correlation are treated within the local-spin-density or gradient-corrected approximations. The basis functions and the electron density are projected on a real-space grid in order to calculate the Hartree and exchange–correlation potentials and matrix elements. We substitute the customary diagonalization procedure by the minimization of a modified energy functional, which gives orthogonal wave functions and the same energy and density as the Kohn–Sham energy functional, without the need of an explicit orthogonalization. The additional restriction to a finite range for the electron wave functions allows the computational effort (time and memory) to increase only linearly with the size of the system. Forces and stresses are also calculated efficiently and accurately, allowing structural relaxation and molecular dynamics simulations. We present test calculations beginning with small molecules and ending with a piece of DNA. Using double-z, polarized bases, geometries within 1\% of experiments are obtained. © 1997 John Wiley \& Sons, Inc. Int J Quant Chem 65: 453–461, 1997},
author = {S\'{a}nchez-Portal, Daniel and Ordej\'{o}n, Pablo and Artacho, Emilio and Soler, Jos\'{e} M},
doi = {10.1002/(SICI)1097-461X(1997)65:5<453::AID-QUA9>3.0.CO;2-V},
journal = {Int. J. Quantum Chem.},
pages = {453--461},
title = {{Density-functional method for very large systems with LCAO basis sets}},
volume = {65},
year = {1997}
}
@article{Sanvito1999,
abstract = {Using a scattering technique based on a parametrized linear combination of atomic orbitals Hamiltonian, we calculate the ballistic quantum conductance of multi-wall carbon nanotubes. We find that inter-wall interactions not only block some of the quantum conductance channels, but also redistribute the current non-uniformly over the individual tubes across the structure. Our results provide a natural explanation for the unexpected integer and non-integer conductance values reported for multi-wall nanotubes in Ref.[S. Frank, P. Poncharal, Z.L. Wang, and W. A. de Heer, Science 280, 1744 (1998)]},
author = {Sanvito, S. and Kwon, Y-K. and Tom\'{a}nek, D. and Lambert, C. J.},
doi = {10.1103/PhysRevLett.84.1974},
journal = {Phys. Rev. Lett.},
pages = {1974--1977},
title = {{Fractional quantum conductance in carbon nanotubes}},
volume = {84},
year = {1999}
}
@article{Shan2005,
abstract = {We perform first principles calculations on work functions of single$\backslash$nwall carbon nanotubes, which can be divided into two classes according$\backslash$nto tube diameter (D). For class I tubes (D>1 nm), work functions$\backslash$nlie within a narrow distribution (∼0.1 eV) and show no significant$\backslash$nchirality or diameter dependence. For class II tubes (D<1 nm), work$\backslash$nfunctions show substantial changes, with armchair tubes decreasing$\backslash$nmonotonically with diameter, while zigzag tubes show the opposite$\backslash$ntrend. Surface dipoles and hybridization effects are shown to be$\backslash$nresponsible for the observed work function change.},
author = {Shan, Bin and Cho, Kyeongjae},
doi = {10.1103/PhysRevLett.94.236602},
journal = {Phys. Rev. B},
pages = {236602--1\~{}4},
title = {{First Principles Study of Work Functions of Single Wall Carbon Nanotubes}},
volume = {94},
year = {2005}
}
@article{Song2008,
abstract = {Glycines are spontaneously adsorbed to form into self-assembled nanoclutsers on single-walled carbon nanotubes (SWNTs). After formation of glycine nanoclusters on SWNTs, the field effect transistor (FET) devices show selective sensing ability to alcohols, such as isopropyl alcohol (IPA), methanol, and ethanol. Upon the adsorption of alcohol, the glycine-coated SWNT-FET devices exhibit pseudo-metallic transport behaviors, whereas the original and glycine-coated devices display conventional p-type transport characteristics. Computational studies support that the gate field screening effect induced by instantly formed glycine- alcohol pair layers seems to be responsible for the pseudo-metallic transport behavior.},
author = {Song, Hyun Jae and Lee, Yoonmi and Jiang, Tao and Kussow, Adil Gerai and Lee, Minbaek and Hong, Seunghun and Kwon, Young-Kyun and Choi, Hee Cheul},
doi = {10.1021/jp077049c},
journal = {J. Phys. Chem. C},
pages = {629--634},
title = {{Self-clusterized glycines on single-walled carbon nanotubes for alcohol sensing}},
volume = {112},
year = {2008}
}
@article{Tans1998,
abstract = {The use of individual molecules as functional electronic devices was first proposed in the 1970s (ref. 1). Since then, molecular electronics2,3 has attracted much interest, particularly because it could lead to conceptually new miniaturization strategies in the electronics and computer industry. The realization of single- molecule devices has remained challenging, largely owing todifficulties in achieving electrical contact to individual molecules. Recent advances in nanotechnology, however, have resulted in electrical measurements on single molecules4–7 . Here we report the fabrication of a field-effect transistor—a three-terminal switching device—that consists of one semiconducting8–10 single-wall carbon nanotube11,12 By applying a voltage to a gate electrode, the nanotube can be switched from a conducting to an insulating state. We have previously reported5 similar behaviour for a metallic single-wall carbon nanotube operated at extremely low temperatures. The present device, in contrast, operates at roomtemperature, thereby meeting an important requirement for potential practical applications. Electrical measurements on the nanotube tran- sistor indicate that its operation characteristics can be qualita- tively described by the semiclassical band-bending models currently used for traditional semiconductor devices. The fabrication of the three-terminal switching device at the level of a single molecule represents an important step towards molecular electronics.},
author = {Tans, Sander J S.J. and Verschueren, Alwin R M A.R.M. and Dekker, Cees},
journal = {Nature},
pages = {49--52},
title = {{Room-temperature transistor based on a single carbon nanotube}},
volume = {393},
year = {1998}
}
@article{Thostenson2005,
author = {Thostenson, E and Li, C and Chou, T},
doi = {10.1016/j.compscitech.2004.11.003},
journal = {Compos. Sci. Technol.},
pages = {491--516},
title = {{Nanocomposites in context}},
volume = {65},
year = {2005}
}
@article{Tombler2000,
abstract = {The effects of mechanical deformation on the electrical properties of carbon nanotubes are of interest given the practical potential of nanotubes in electromechanical devices, and they have been studied using both theoretical and experimental approaches. One recent experiment used the tip of an atomic force microscope (AFM) to manipulate multi-walled nanotubes, revealing that changes in the sample resistance were small unless the nanotubes fractured or the metal-tube contacts were perturbed. But it remains unclear how mechanical deformation affects the intrinsic electrical properties of nanotubes. Here we report an experimental and theoretical elucidation of the electromechanical characteristics of individual single-walled carbon nanotubes (SWNTs) under local-probe manipulation. We use AFM tips to deflect suspended SWNTs reversibly, without changing the contact resistance; in situ electrical measurements reveal that the conductance of an SWNT sample can be reduced by two orders of magnitude when deformed by an AFM tip. Our tight-binding simulations indicate that this effect is owing to the formation of local sp3 bonds caused by the mechanical pushing action of the tip.},
author = {Tombler, Tw and Zhou, C and Alexseyev, L and Kong, J and Dai, H and Liu, L and Jayanthi, Cs and Tang, M and Wu, Sy},
doi = {10.1038/35015519},
file = {:C$\backslash$:/Users/yjlee/Desktop/CNT논문 Ref/youngref/405769a0.pdf:pdf},
journal = {Nature},
number = {1993},
pages = {769--72},
title = {{Reversible electromechanical characteristics of carbon nanotubes under local-probe manipulation}},
volume = {405},
year = {2000}
}
@article{Tournus2005,
abstract = {The -stacking interaction between various planar organic molecules is investigated within the framework of ab initio calculations. The adsorption of these molecules on the sidewall of the cylindrical carbon structure induces a small binding energy compared to conventional covalent functionalization. Such a weak interaction is found to be only physisorption and leads to minor and predictable modifications of the electronic structure. These changes in the electronic behavior of the host carbon nanotube are ruled by the relative positions of the molecular levels of the isolated molecule and both the valence and conduction bands of the perfect tube.},
author = {Tournus, F. and Latil, S. and Heggie, M. and Charlier, J.-C.},
doi = {10.1103/PhysRevB.72.075431},
journal = {Phys. Rev. B},
pages = {075431--1\~{}5},
title = {$\pi$-stacking interaction between carbon nanotubes and organic molecules},
volume = {72},
year = {2005}
}
@article{Valavala2008,
abstract = {First-principles density-functional calculations of the electronic structure, energy band gaps ?Eg?, and strain- induced band gap changes in moderate-gap single-walled ?n,0? carbon nanotubes ?SWNTs? are presented. It is confirmed that ?n,0? SWNTs fall into two classes depending upon n mod 3=1 or 2. Eg is always lower for “mod 1” than for “mod 2” SWNTs of similar diameter. For n?10, strong curvature effects dominate Eg; from n=10 to 17, the Eg oscillations, amplified due to ? ? - mixing, decrease and can be explained very well with a tight-binding model which includes trigonal warping. Under strain, the two families of semiconducting SWNTs are distinguished by equal and opposite energy shifts for these gaps. For ?10,0? and ?20,0? tubes, the potential surface and band gap changes are explored up to approximately ?6\% strain or compression. For each strain value, full internal geometry relaxation is allowed. The calculated band gap changes are ??115?10? meV per 1\% strain, positive for the mod 1 and negative for the mod 2 family, about 10\% larger than the tight-binding result of ?97 meV and twice as large as the shift predicted from a tight-binding model that includes internal sublattice relaxation.},
author = {Valavala, Pavan K. and Banyai, Douglas and Seel, Max and Pati, Ranjit},
doi = {10.1103/PhysRevB.78.235430},
journal = {Phys. Rev. B},
pages = {1--6},
title = {{Self-consistent calculations of strain-induced band gap changes in semiconducting (n,0) carbon nanotubes}},
volume = {78},
year = {2008}
}
@article{Van2000,
abstract = {The first all-electron ab initio study of Young's modulus and Poisson ratio for a no. of closed single-walled nanotubes is presented. At the Hartree-Fock 6-31G* level, the results obtained compare well with exptl. as well as previous theor. studies, predicting a Young's modulus higher than 1 TPa. The calcd. Young's modulus for a graphene layer is found to be smaller than for its (5,5)-nanotube counterpart. [on SciFinder(R)]},
author = {Van, Lier G and Van, Alsenoy C and Van, Doren V and Geerlings, P},
doi = {10.1016/s0009-2614(00)00764-8},
file = {:C$\backslash$:/Users/yjlee/Desktop/CNT논문 Ref/youngref/1-s2.0-S0009261400007648-main.pdf:pdf},
journal = {Chem. Phys. Lett.},
number = {August},
pages = {181--185},
title = {{Ab initio study of the elastic properties of single-walled carbon nanotubes and graphene}},
volume = {326},
year = {2000}
}
@article{Yin2000,
abstract = {Grand canonical ensemble Monte Carlo (GCEMC) molecular simulations of hydrogen storage at 298 and 77 Kin triangular arrays of single wall carbon nanotubes (SWCNT) and in slit pores (modeling activated carbons) were performed. At 298 K the US DOE target gravimetric hydrogen storage capacity (6.5 wt \%) is reached at 160 bar for optimally configured arrays of open SWCNT of wide diameter, but the equivalent volumetric capacity is similar to 40\% of the DOE target [695 (STP)v/v]. For slit pores at 298 K the optimal volumetric capacity is similar to 20\% of the target. Simulations for 77 K and 70 bar indicate that triangular arrays of open and closed SWCNT of various diameters in a wide range of configurations exceed the DOE gravimetric target. A capacity of 33 wt \% is found for arrays of narrow, open, or closed SWCNT that are widely spaced. Here, adsorption occurs entirely in the interstitial space between the nanotubes. Volumetric capacities close to the DOE target are found for arrays of narrow, open or closed SWCNT with a range of interstitial spacings. The maximum volumetric capacities for simulations with slit pores at 77 K and 70 bar are similar to 73\% of the DOE target for a range of pore widths. Capacities from simulations for nanotubes and slit pores at 298 and 77 K are in reasonable agreement with experimentally measured capacities. It is concluded that the potential of carbon nanotubes for storage of hydrogen is superior to that of activated carbons.},
author = {Yin, Y F and Mays, T and McEnaney, B},
doi = {10.1021/la000900t},
journal = {Langmuir},
number = {16},
pages = {10521--10527},
title = {{Molecular simulations of hydrogen storage in carbon nanotube arrays}},
volume = {16},
year = {2000}
}
@article{Yu2000,
abstract = {The mechanical response of 15 single wall carbon nanotube (SWCNT) ropes under tensile load was measured. For 8 of these ropes strain data were obtained and they broke at strain values of 5.3\% or lower. The force-strain data are well fit by a model that assumes the load is carried by the SWCNTs on the perimeter of each rope. This model provides an average breaking strength of SWCNTs on the perimeter of each rope; the 15 values range from 13 to 52 GPa (mean 30 GPa). Based on the same model the 8 average Young's modulus values determined range from 320 to 1470 GPa (mean 1002 GPa).},
author = {Yu, Min-Feng and Files, Bradley S and Arepalli, Sivaram and Ruoff, Rodney S},
doi = {10.1103/PhysRevLett.84.5552},
journal = {Phys. Rev. Lett.},
pages = {5552--5555},
title = {{Tensile Loading of Ropes of Single Wall Carbon Nanotubes and their Mechanical Properties}},
volume = {84},
year = {2000}
}
@article{Yao1999,
abstract = {The ultimate device miniaturization would be to use individual molecules as functional devices. Single-wall carbon nanotubes (SWNTs) are promising candidates for achieving this: depending on their diameter and chirality, they are either one-dimensional metals or semiconductors1,2. Single-electron transistors employ- ing metallic nanotubes3,4 and ®eld-effect transistors employing semiconducting nanotubes5 have been demonstrated. Intramol- ecular devices have also been proposed which should display a range of other device functions6±11. For example, by introducing a pentagon and a heptagon into the hexagonal carbon lattice, two tube segments with different atomic and electronic structures can be seamlessly fused together to create intramolecular metal± metal, metal ± semiconductor, or semiconductor ± semiconductor junctions. Here we report electrical transport measurements on SWNTs with intramolecular junctions. We ®nd that a metal± semiconductor junction behaves like a rectifying diode with nonlinear transport characteristics that are strongly asymmetric with respect to bias polarity. In the case of a metal±metal junction, the conductance appears to be strongly suppressed and it displays a power-law dependence on temperatures and applied voltage, consistent with tunnelling between the ends of two Luttinger liquids. Our results emphasize the need to consider screening and electron interactions when designing and model- ling molecular devices. Realization of carbon-based molecular electronics will require future efforts in the controlled production of these intramolecular nanotube junctions.},
author = {Z, Yao. and Postma, H.W.Ch. and L, Balents and C, Dekker},
doi = {10.1088/0957-4484/18/39/395205},
journal = {Nature},
number = {November},
pages = {273--276},
title = {{Carbon Nanotube Intramolecular Junctions}},
volume = {402},
year = {1999}
}
@article{Zhou2000,
abstract = {Modulation doping of a semiconducting single-walled carbon nanotube along its length leads to an intramolecular wire electronic device. The nanotube is doped n-type for half of its length and p-type for the other half. Electrostatic gating can tune the system into p-n junctions, causing it to exhibit rectifying characteristics or negative differential conductance. The system can also be tuned into n-type, exhibiting single-electron charging and negative differential conductance at low temperatures. The low-temperature behavior is manifested by a quantum dot formed by chemical inhomogeneity along the tube.},
author = {Zhou, C and Kong, J and Yenilmez, E and Dai, H},
doi = {10.1126/science.290.5496.1552},
journal = {Science (80-. ).},
pages = {1552--1555},
title = {{Modulated chemical doping of individual carbon nanotubes.}},
volume = {290},
year = {2000}
}
@article{Zolyomi2004,
author = {Z\'{o}lyomi, V. and K\"{u}rti, J.},
doi = {10.1103/PhysRevB.70.085403},
journal = {Phys. Rev. B},
pages = {085403},
title = {{First-principles calculations for the electronic band structures of small diameter single-wall carbon nanotubes}},
volume = {70},
year = {2004}
}
@article{Zhang2009,
author = {Zhang, Yuanjian and Thomas, Arne and Antonietti, Markus and Wang, Xinchen},
doi = {10.1021/ja808329f},
journal = {J. Am. Chem. Soc.},
pages = {50--51},
title = {{Activation of Carbon Nitride Solids by Protonation: Morphology Changes, Enhanced Ionic Conductivity, and Photoconduction Experiments}},
volume = {131},
year = {2009}
}
@article{Zambov2000,
author = {Zambov, By Ludmil M and Popov, Cyril and Abedinov, Nikolai and Plass, Martin F and Kulisch, Wilhelm and Gotszalk, Teodor and Grabiec, Peter and Rangelow, Ivo W and Kassing, Rainer},
file = {:E$\backslash$:/Aspirin/656\_ftp.pdf:pdf},
journal = {Adv. Mater.},
number = {9},
pages = {656--660},
title = {{Gas-sensitive properties of nitrogen-rich carbon nitride films}},
volume = {12},
year = {2000}
}
@article{Li1995,
author = {Li, Dong and Chu, Xi and Cheng, Shang-Cong and Lin, Xi-Wei and Dravid, Vinayak P. and Chung, Yip-Wah and Wong, Ming-Show and Sproul, William D.},
doi = {10.1063/1.114667},
file = {:E$\backslash$:/Aspirin/coating.pdf:pdf},
journal = {Appl. Phys. Lett.},
number = {2},
pages = {203},
title = {{Synthesis of superhard carbon nitride composite coatings}},
volume = {67},
year = {1995}
}
\end{document}